# Milagrito, a TeV air-shower array

R. Atkins,[1] W. Benbow,[2] D. Berley,[3,a] M.-L. Chen,[3,b] D. G. Coyne,[2] R. S. Delay,[4] B. L. Dingus,[1] D. E. Dorfan,[2] R. W. Ellsworth,[5] C. Espinoza,[6] D. Evans,[3] A. Falcone,[7] L. Fleysher,[8] R. Fleysher,[8] G. Gisler,[6] J. A. Goodman,[3] T. J. Haines,[6] C. M. Hoffman*,[6] S. Hugenberger,[4] L. A. Kelley,[2] S. Klein,[2,c] I. Leonor,[4] J. Macri,[7] M. McConnell,[7] J. F. McCullough,[2] J. E. McEnery,[1] R. S. Miller,[6,d] A. I. Mincer,[8] M. F. Morales,[2] M. M. Murray,[6] P. Nemethy,[8] G. Paliaga,[2] J. M. Ryan,[7] M. Schneider,[2] B. Shen,[9] A. Shoup,[4] G. Sinnis,[6] A. J. Smith,[9,e] G. W. Sullivan,[3] T. N. Thompson,[6] O. T. Tumer,[9] K. Wang,[9] M. O. Wascko,[9] S. Westerhoff,[2] D. A. Williams,[2] T. Yang,[2] G. B. Yodh[4]   (The Milagro Collaboration)

[1] University of Utah, Salt Lake City, UT 84112
[2] University of California, Santa Cruz, CA 95064
[3] University of Maryland, College Park, MD 20742
[4] University of California, Irvine, CA 92717
[5] George Mason University, Fairfax, VA 22030
[6] Los Alamos National Laboratory, Los Alamos, NM 87545
[7] University of New Hampshire, Durham, NH 03824-3525
[8] New York University, New York, NY 10003
[9] University of California, Riverside, CA 92521
[a] Permanent Address: National Science Foundation, Arlington, VA 22230
[b] Now at Brookhaven National Laboratory, Upton, NY 11973
[c] Now at Lawrence Berkeley National Laboratory, Berkeley, CA 94720
[d] Now at University of New Hampshire, Durham, NH 03824-3525
[e] Now at University of Maryland, College Park, MD 20742

Abstract

Milagrito, a large, covered water-Cherenkov detector, was the world's first air-shower-particle detector sensitive to cosmic gamma rays below 1 TeV. It served as a prototype for the Milagro detector and operated from February 1997 to May 1998. This paper gives a description of Milagrito, a summary of the operating experience, and early results that demonstrate the capabilities of this technique.

# 1. Introduction

High-energy gamma-ray astronomy allows a unique study of non-thermal, energetic acceleration processes in the universe. Promising objects for study include pulsars, active galaxies, supernova remnants, gamma-ray bursts (GRBs), and black holes. Other potential sources include more esoteric phenomena such as evaporating primordial black holes, topological defects, and neutralino annihilation. Reviews of the techniques, science, and recent results in high-energy gamma-ray astronomy are available [1,2].

Cosmic gamma rays up to a few GeV can be directly detected with satellite-based detectors, such as EGRET [3]. At higher energies, the gamma-ray flux from even the brightest source is too low to be measured in the relatively small detectors that can be placed in satellites: thus ground-based techniques are used. High-energy gamma rays interact high in the atmosphere producing a cascade of particles called an extensive air shower (EAS). Ground-based gamma-ray telescopes detect the products of an EAS that survive to ground level.

After many years of perfecting the technique, atmospheric Cherenkov telescopes (ACT) have been successfully employed in the energy region from ~400 GeV – 10 TeV. ACTs detect Cherenkov light produced in the atmosphere by the ultra-relativistic particles in the EAS. ACTs have detected emission from a number of sources including three plerions, the Crab, Vela, and PSR1706-44, and four active galaxies, Markarian 421, Markarian 501, 1ES2344+514, and PKS2155-304. These observations have greatly improved our understanding of the acceleration mechanisms in these sources. There are many efforts underway to improve ACTs, primarily to increase the flux sensitivity [4,5] and to lower the energy threshold [6,7,8].

Extensive air-shower particle detector arrays (EAS arrays) have been used in the energy regime above ~50 TeV. An EAS array typically consists of a sparse array of counters that detect a small fraction of the charged shower particles that reach ground level. Unfortunately, no convincing evidence for steady gamma-ray emission above 50 TeV from any source has been obtained with these detectors.

The excellent angular resolution and sensitivity of ACTs make them ideal to study emission from steady and highly variable sources. However, ACTs can only be used on clear, dark nights, and can only view one source at a time (and only during that part of the year when that source is in the night sky). Thus they are not well suited to perform an all-sky survey or to search for episodic emission from a known source or from a source at an unknown direction (such as from a GRB). On the other hand, an EAS array can operate 24 hours per day, regardless of weather, and can observe the entire overhead sky; an EAS array is able to observe every source in its field of view every day of the year. It would be worthwhile to perform high-duty-factor, all-sky observations in the energy region presently being studied by ACTs. Consequently there are a number of efforts to lower the energy threshold of EAS arrays, including siting the detector at an extremely high altitude [9], and using an instrument that can detect a large fraction of the shower particles that reach the ground within the detector area. The Milagrito detector, described in this paper, follows the second approach.

In this paper we describe the design, construction, and operation of the Milagrito detector. Milagrito was built as a prototype for Milagro [10]. However, it was also a fully functioning detector that took data from February 1997 to May 1998. With Milagrito we studied specific design questions for Milagro, learned how to operate a remote detector, took data to get a first glimpse at the astrophysics that can be done with such a device, and developed reconstruction and analysis techniques.

## 2. Description of the Milagrito Detector
### a) The Concept

The goal of the Milagro project is to build a detector sensitive to cosmic gamma rays around 1 TeV while maintaining the all-sky, high duty-factor capabilities of an EAS array. A typical EAS array has scintillation counters covering <1% of the total area of the array. Thus, at best, only a small fraction of the shower particles surviving to ground level are detected, which results in a high threshold energy. This is compounded by the fact that showers at ground level contain many more gamma rays than charged particles and scintillation counters do not detect gamma rays with high efficiency.

Milagro uses photomultiplier tubes (PMTs) deployed under ~1.5 m of water to detect the Cherenkov radiation produced in the water by relativistic charged shower particles. Because water is inexpensive and the Cherenkov cone spreads out the light, one is able to construct a large instrument that can detect nearly every charged shower particle falling within its area. Furthermore, the plentiful gamma rays convert to electron-positron pairs (or electrons via Compton scattering) that, in turn, produce Cherenkov radiation that can be detected. Consequently, Milagro has a very low energy threshold for an EAS array.

As in a conventional EAS array, the direction of the primary gamma ray is reconstructed by measuring the relative times at which the individual PMTs are struck by light produced by particles in the shower front. This is illustrated in Figure 1.

The amount of water above the PMTs should be large enough for nearly all of the photons to convert into electrons and positrons, but not so thick that the light produced at the top is absorbed before it reaches the PMT. The width of the observed timing distributions increases as the depth of water above the PMTs increases. This is caused by two effects: the difference in the arrival time of Cherenkov light produced at the top of the layer and that produced by a pair created near the PMT increases, and the geometrical area viewed by each PMT increases. Countering these effects is the statistical gain from collecting light from a larger fraction of particles as the water depth above the PMTs increases. Thus the angular resolution and the energy threshold depend on the water depth. The optimal depth gives the best sensitivity to a signal from a gamma-ray source. From Monte Carlo simulations one expects the optimal depth to be a 1-2 meters. Milagrito data were taken with several water depths to study how its performance changes with depth.

Milagrito had 228 PMTs located in a plane on a 2.8 m x 2.8 m grid near the bottom of the pond. Data were taken with 1, 1.5 and 2 meters of water above the tubes.

### b) The Pond

Milagrito was built in a pre-existing high-altitude water reservoir in the Jemez Mountains near Los Alamos, New Mexico. The site, which is at an altitude of 2650 m above sea level (750 g/cm$^2$ atmospheric overburden), was formerly used for the Hot Dry Rock geothermal energy project. The 21-million-liter pond is rectangular with dimensions of 60 m x 80 m at the surface and sloping sides that lead to a bottom of 30 m x 50 m at a depth of 7.5 m (see Figure 2). The site also has a fresh-water well and pumps to deliver water to the pond.

The pond became available to the Milagro project in 1995. A new high-density polypropylene liner and cover were installed [11]. This 1 mm-thick material has two black layers of polypropylene with an internal polyester scrim. The opacity of the cover was a chief concern. The requirement to detect single photons during the day demands that the cover transmit essentially no light. The two black layers guarantee that this requirement is satisfied. Straps sewn into the cover can be tightened to take up the slack in the cover when the pond is full. The cover can be inflated so that people may enter and work on the detector (see below). The straps are also used to keep the cover taut during inflation. These straps are plainly visible in the aerial view of the Milagro site (Figure 2). The top of the cover was subsequently painted with a highly reflective roofing paint, reducing the temperature under the inflated cover by about 10$^o$ C to 35$^o$ C on sunny days.

The photomultiplier tubes used in Milagrito were anchored by 1.6 mm-diameter Kevlar strings to a grid of weighted PVC pipe, 7.5 cm in diameter. The PVC pipes were arranged in a square grid with 2.8 m spacing on the bottom of the pond. The pipe spacing on the sloping sides of the pond was arranged so that the entire pond appears as a continuous grid of uniform spacing when viewed vertically from above. The length of each Kevlar string was calculated for each PMT based on a survey of the contour of the pond bottom to an accuracy of ~1 cm, so that all PMTs would lie in a horizontal plane when they float upwards. The pipes were filled with wet sand. Each filled 2.8 m-long pipe weighed ~35 kg. On the sides of the pond, only the pipes going up the sides were filled with sand, as the unsupported horizontal cross pieces would sag if they were weighted. Each of these vertical pipes was 4 m long and weighed ~45 kg.

The complete set of PMT string attachment points was surveyed after the grid was installed. The relative accuracy of the survey of each grid corner is ±3 cm horizontally and ±1 cm vertically. The orientation of the grid with respect to true north was determined using the location of the sun and a GPS clock to an accuracy of ±0.02$^o$; the direction of the zenith is also known to ±0.02$^o$.

### c) The Physical Plant

A utility building was constructed next to the pond to house the water filtration and recirculation system, the inflation system for the pond cover, and a patch panel for the PMT cables. The building is connected to the pond with two polypropylene tunnels: one for air from the cover inflation system and the other for the cables and the water pipes. The cables emerge from the pond and connect to an interface patch panel, then continue via underground conduit to an electronics trailer located near the utility building. The electronics trailer houses the electronics, the data acquisition system, computers, and communications equipment.

The cover can be inflated using a system of fans and filters to blow 135 m$^3$/min of clean air into the pond. The pressure under the cover is raised to an overpressure of 25-50 Pa, depending on wind conditions. The cover was stabilized by 20 polyester straps connected to hand winches on the sides of the pond. It was found that this arrangement was not sufficient to stabilize the cover under moderate winds. An improved cover restraint system was installed prior to the upgrade from Milagrito to Milagro.

The Milagrito water was drawn from the fresh water well located on the site. The water filtration system consisted of a media filter, a water softener, a carbon filter, a 1-µm filter, a UV lamp, and a 0.2-µm filter. The fill rate was 300 liters/min. During operations the total amount of water in the pond was between 3,000,000 and 4,400,000 liters depending upon the depth. A pump located at the bottom of the pond was used to continuously recirculate the water at 725 liters/min during normal operations. The media filter and softener were bypassed during recirculation. Figure 3 shows the attenuation length of the water at a wavelength of 350 nm as a function of date: the last four points are for Milagro (see Section 11 below). The errors shown are those of the measurement itself and do not include any contribution from the processes of collecting the water and shipping it to California, where the measurements were made. It should be noted that the measured attenuation length includes both absorption and scattering of the light.

Milagrito is located in one of the most lightning-prone areas in the US. The mean waiting time for a strike within the 50,000 m$^2$ site is about 1 month, peaking in the early summer. To protect the experiment, a 12,500 m$^2$ Faraday cage was erected over the entire pond and adjoining auxiliary buildings. The philosophy of the protection system is to intercept (not avoid) lightning strokes and to shunt their current to ground, keeping the voltage gradients low within the environs of the experiment. The cage is made of a mesh of 1/0 and 3/0 stranded hard-copper wire on a rectangular grid spacing of 5.2 m x 21 m and designed to withstand the highest lightning current expected over ~20 years. The wires droop ~3.5 m from the 12 m-high telephone poles, allowing the cover to be inflated to a height of about 6 m above grade. The system is able to withstand the tension incurred by reasonable ice loads in winter. The wire ends are anchored and bonded to the steel chain-link fence surrounding the pond. A 1.8 m copper lightning rod with a rounded top is affixed to the top of each pole. Each lightning rod is connected to a buried counterpoise grid that connects all of the poles. Underground wires connect this grid to the peripheral chain-link fence.

The site is fed with three-phase 240 V power from the local rural electric company. A motor-generator (MG) set is used to condition this power for use by the electronics. The MG set must be manually reset after power outages. The site has several telephone lines and network access via a T1 line. The T1 line was used for remote monitoring and some control of the apparatus.

**3. Photomultiplier tubes**

The heart of the Milagrito detector is the array of photomultiplier tubes deployed under water to detect air-shower particles reaching the ground. These PMTs must measure the arrival time and intensity of the air shower. Important PMT characteristics in this application include:

      a) Good time resolution,
      b) Good charge resolution including a resolved 1-photoelectron peak,
      c) Good charge linearity over a wide dynamic range,
      d) High efficiency and large photocathode area,
      e) Relative absence of significant prepulsing and afterpulsing, and
      f) Relative insensitivity to the effects of the geomagnetic field.

The 20 cm-diameter Hamamatsu 10-stage R5912SEL PMT was selected. Figure 4 shows the measured PMT time resolution for single-photoelectron (PE) pulses. Figure 5 shows the single photoelectron charge resolution for this PMT.

The pinout of each PMT is encapsulated in a waterproof PVC housing to keep it and the PMT base dry. The PMT base was coated in a silicone conformal coating to protect the components from humidity. In the initial design, the housings were attached to the glass of the PMT with a rigid epoxy. However, the contraction of the PVC at low temperatures (below $\sim 0^o$ C) produced sufficient stress to crack the PMT glass. Consequently, the design was changed and Dow Corning RTV 3145 was used to attach the PMT to its encapsulation. All PMTs with housings passed a pressure test for water leakage (120 kPa for 24 hours) prior to installation.

The PMTs operate at positive high voltage so that the photocathode is at the same potential as the surrounding water. A passive base with a total resistance of 20 M$\Omega$ is used. A single 75$\Omega$ coaxial cable (RG59; Belden Part #YR29304), 137 m long, supplies the high voltage to the PMT and carries the PMT signal to the electronics. This cable has a watertight coaxial connector (W. W. Fischer, Inc. Part #SE103 A002/6.2) that connects to a watertight coaxial bulkhead connector (W. W. Fischer, Inc. Part #DBEE 103 A002) mounted on the side of the PVC housing. Great care was required to properly assemble the connector on the cable.

The linearity of several PMTs and bases have been tested using a 337-nm-wavelength nitrogen laser with ~0.3 ns pulse width and a set of calibrated neutral density filters. The results, shown in Figure 6 for five PMTs operating at a gain of ~2 x 10$^7$, show that the PMTs are linear up to ~75 photoelectrons.

**4. Electronics**
The main function of the electronics is to provide timing and pulse height information from each PMT channel. In addition, a trigger decision must be made to select showers for digitization and the accurate time of arrival of the selected showers must be recorded. Custom-made electronics boards distributed high voltage to the PMTs and processed the PMT signals prior to readout using commercial FASTBUS modules.

The pulse height for each PMT was measured using the time-over-threshold (TOT) technique. The requirements of good timing information independent of PMT prepulsing and a large dynamic range on the pulse height measurement led to a dual-threshold system. Both timing and TOT information were determined for two different discriminator thresholds: low threshold (~1/4 PE) and high threshold (~5 PE). In addition to these functions the front-end boards also provided triggering and monitoring information.

An "analog board" did the initial signal processing of the PMT signals. Each board distributed the high voltage to PMTs in two groups of 8 channels each; the two groups could be connected together if desired so that one external HV channel could supply 16 PMTs. One resistor per channel allowed voltage adjustment of individual channels. A single RG-59 cable provided high voltage to each PMT and carried the signal from the PMT. The signal was AC coupled to the amplifier inputs by a high voltage capacitor on the analog board. The analog signal from each PMT was split and passed through two amplifiers with different gains. The amplified signals were then integrated on capacitors, which discharged with a time constant of 100 ns. This value of the time constant was chosen to minimize the effect of late light on the measurement of the TOT. The output from the higher-gain amplifier was the input to the low-threshold discriminator. The amplifier gain and discriminator threshold were set so that a PMT signal of ~1/4 PE would fire the discriminator. The output of the lower-gain amplifier was split into two paths. One output went to a passive delay-line chip, providing a delayed analog pulse that could be integrated with an external analog-to-digital converter (ADC). The other output went to a discriminator with a threshold of ~5 PE, the high threshold. Both discriminators generated TOT pulses. The discriminator outputs were routed to a digital board that performed the digital signal processing.

The digital boards multiplexed the low- and high-threshold discriminator signals (see Figure 7) and provided triggering and monitoring information. Each PMT signal crossing a discriminator threshold generated a fixed duration, 300 ns pulse of 25 mV amplitude. The triggering information is simply the analog sum of these fixed duration pulses, similar to a multiplicity level output. Multiplicity information was independently provided for both low and high-threshold discriminators. For Milagrito, only the low-threshold multiplicity information was used to form a simple multiplicity trigger. Figure 8 shows the trigger rate as a function of the number of PMTs required for a water depth of 1 m. Milagrito data were taken with a requirement of >100 PMTs within the 300 ns coincidence window.

CAMAC scalers were used to monitor the performance of the PMTs and to extend the detector sensitivity to lower-energy showers (see section 9). The low-threshold outputs from sets of 4 OR'ed nonadjacent PMTs and the high-threshold outputs from sets of 16 OR'ed PMTs were counted every second. In addition, the low-threshold outputs from individual PMTs could be counted via a programmable mask.

The Milagrito timing and pulse height information were encoded as a series of edges, as illustrated in Figure 7. These data were digitized in three LeCroy 1887 FASTBUS TDC modules. Each module has 96 channels that can each record up to 16 edges per event with 0.5-ns resolution, and 8 event buffers. The time of each event was recorded by latching the output of a GPS clock in a Struck Model 137/2 FASTBUS Latch. In addition, a LeCroy 1881M FASTBUS ADC was used to digitize the analog outputs of a subset of the PMTs for calibration purposes; the ADC was manually cycled through all the channels.

**5. Data Acquisition**
The data were read out after digitization with a FASTBUS Smart Crate Controller (FSCC) [12]. The FSCC transferred the data to a pair of dual-ported VME memory modules via a smart controller (Access Dynamics DM115/DC2). The use of two memory boards allowed for the simultaneous reading and writing of data. The controller wrote the data into the memory boards over the VSB bus. A Silicon Graphics (SGI) Challenge L multi-CPU computer read the data from the memory boards over the VME bus. Because the DM115/DC2 board did not fit in the internal VME crate of the SGI, an external VME crate was used with a BIT-3 bus repeater (Model 418/418-50) connecting the two VME crates. Milagrito typically operated at a trigger rate of 300-400 Hz with less than 0.5% dead time. The data were written to disk before being archived to DLT tape.

The experiment was controlled by commands issued from the SGI computer. The computer communicated with the FSCC via the Ethernet port on the FSCC. From the SGI one could initialize the FASTBUS modules and prepare the system for data taking. Once data taking began the online system monitored error rates and requested resets of the FASTBUS system as necessary. In cases that required human intervention, an alert was sent and the person on shift received a pager message. The system could be monitored and controlled remotely via the web.

The event processing was performed by several semi-autonomous routines. The event calibration was performed by the routine that read the data from the VME memories. These data were put into a shared memory location where another routine would access them and perform the event reconstruction. The reconstructed data were deposited into the same shared memory region and another routine accessed it and wrote it out to disk. In all cases the original time ordering of the events was maintained. Throughout the operation of Milagrito, the raw data as well as the results of the event reconstruction were saved so that the event reconstruction could be repeated offline.

Milagrito had an independent, computer-based system to monitor the status of many hardware components. This system (called the Environment Monitoring System, EMS) used a PC running Linux to monitor and record information from a number of transducers and other hardware. This information included:

- Weather information—air pressure, temperature (both inside and outside the electronics trailer), humidity, wind speed and direction, and accumulated rain.
- Water temperature inside the pond, both at the pond bottom and the surface.
- Pond depth (which includes the amount of accumulated water on the cover).

- Water recirculation system status, including the water temperature, several flow rates, the status of the UV light, and the water pressure at seven locations in the system.
- The output voltage and current of each high-voltage channel.
- The temperature at various locations inside the electronics racks.

The EMS also read out the CAMAC scalers that counted PMT rates. All of these quantities were available remotely via a web page that included daily, monthly, and yearly plots of these quantities and a list of problems with individual PMTs (such as abnormally low or high count rates).

## 6. Calibration

The raw PMT data from Milagrito consisted of a series of times and edge senses (leading or trailing). In order to reconstruct the direction and size of the air shower, this raw edge information must be converted to relative arrival times and pulse heights. The correct leading edge must be selected to obtain the relative arrival time. This decision was based on the edge parity and the pulse height. Since the PMTs are susceptible to prepulsing at large pulse heights, the time of the high-threshold crossing was used to determine the arrival time of large pulse height hits. The next step was to correct for electronic slewing, which is slightly different for each PMT. The time-over-threshold measurements (low and/or high threshold) were then converted to numbers of photoelectrons. Then the data were ready for event reconstruction.

The majority of TDC channels recorded two or four edges, depending on the incident light level, as shown in Figure 7. In practice, 69% of all hit TDC channels had two edges and 23% had four edges; the remaining small fraction had from 1 to 16 edges. In channels with four or more edges, a decision had to be made whether the four edges were the result of a single large pulse or two separated smaller pulses. For a single large pulse producing four edges, the separation between successive pairs of edges had a predictable relationship as a result of time constants in the electronics. These relationships were used by an edge-finding code to examine all TDC hits, to discard those that occurred outside the normal time range, and to select the correct leading edge and number of edges (2 or 4) associated with the hit.

A fast, pulsed laser and a system of fiber optics and small diffusing balls were used to determine the bulk of the calibration parameters [13]. The system had to measure the calibration parameters over a wide dynamic range (0.1 PE to >1000 PE). Figure 9 shows a block diagram of the calibration system. The calibration system was designed to measure the electronic slewing for each PMT, measure any fixed timing offsets between PMT channels, measure the conversion between TOT and pulse height, and determine the light level at which prepulsing and saturation were evident in each channel.

A LaserPhotonics Model LN120 nitrogen dye laser, capable of repetition rates of 10-20 Hz, was used. This laser was chosen for its short pulse duration (approximately 300 ps) and high power (0.5 MW per pulse). The selected dye emitted light at 500 nm. This relatively long wavelength was matched to the transmission of the optical fibers to yield the maximum light in the pond. The laser light was focused on a neutral density filter wheel, allowing the light intensity reaching the pond to be varied in a reproducible manner. This filter wheel was mounted on a remote-controlled rotary stage and spanned a range of optical attenuation from 1 (no filtering) to $10^{-4}$. After passing through the filter wheel the laser light was split into 3 equal-intensity beams, with each beam feeding a bank of 10 individual 150m-long fibers. All fibers were connected to a GPIB controlled optical switch. Only 10 of the optical fibers were used in Milagrito. The fibers were tethered from the grid throughout the pond. Each fiber was terminated in a diffusing ball, which floated about 1-meter above the PMTs and pointed downward. The diffusing balls were made by dipping each fiber end into a mixture of titanium dioxide and optical cement.

All channels should have had nearly equal time offsets because the cable lengths and electronics were all nominally the same. However, different PMT operating voltages led to small differences in the time offsets. The determination of the time offsets required determining the location of each diffusing ball relative to the PMT locations, the relative time delays in the fiber optics, the effective speed of light in the water, and the observed time differences between PMTs illuminated by the same fiber. The position of each diffusing ball, the difference in fiber lengths, and the effective speed of light in water were determined by comparing the relative time difference from pairs of fibers observed by a common set of PMTs. The measured speed of light in water is $22.1\pm0.1$ cm/ns, consistent with the expected group velocity of 22.04 cm/ns [14]. The PMT time offsets had a spread of roughly 2 ns.

The time at which a discriminator fired depended upon the size of the input pulse because of the finite rise time of the pulse. The slewing as a function of pulse height was determined for both the low-threshold and high-threshold discriminators for each channel using the laser system by varying the light attenuation of the filter wheel. In addition to the standard electronic slewing (<1 ns at 100 PE and 6-8 ns at 1 PE), there was a ~3 ns per decade charge walk of the arrival time produced by the PMT itself.

An ADC was used to measure the conversion between time-over-threshold and the number of photoelectrons. Simultaneous TDC and ADC information were collected for each PMT channel. First, the ADC spectra were examined to determine the ADC pedestal and the single PE peak for each channel. Then the ADC data (expressed in units of PEs) were plotted vs. the TOT outputs. Figure 10 shows the fractional error in the TOT measurement as a function of pulse height for a typical PMT. The charge resolution of the TOT method is quite good (~10%) over much of the dynamic range. The region near 3 PEs where the resolution approaches 35% is due to the effect of late light on the low-threshold TOT measurement. Once the signal crosses the high-threshold discriminator level, the charge resolution improves.

**7. Event Reconstruction**
Each event was reconstructed to determine the incident direction of the primary particle, the location of the shower core, and the shower size. The reconstruction of an event proceeded in several steps:
1. The location of the shower core was determined.

2. The arrival times at each PMT were corrected for the effect of shower "curvature."
3. The measured arrival time at each PMT was corrected for sampling effects.
4. The appropriate relative weight of each measured arrival time was determined.
5. The direction of the shower plane was determined.

The incident shower-particle swarm is actually not a plane, but is approximately a cone with apex at the shower core. The slope of the cone measured from the shower core is known as the curvature of the shower front. In addition to the curvature, the effect of sampling bias makes the shower appear non-planar: because the shower "plane" actually has a thickness and the arrival time of the first photoelectron is measured by the TDC, the larger the number of PEs, the earlier the measurement (on average).

Once corrections were made for these effects, the arrival times form a plane. The direction of the shower plane was determined using a weighted least squares ($\chi^2$) fit. The $\chi^2$ was minimized with respect to the two independent direction cosines and the arrival time of the shower plane.

Both the sampling correction and the relative weights were determined from the data. This was done by selecting a sample of events that has an unbiased estimate of the shower direction without the use of the above corrections. Such a sample consists of large showers arriving from near the zenith with cores near the center of the pond. An iterative procedure was employed, successively fitting the events and examining the difference between the PMT and shower-plane timing (timing residuals) as a function of pulse height. (In principle the timing residuals depend upon both the pulse and the distance to the core, however in Milagrito the core location was not well determined and therefore the curvature correction was not evident in the data.) The positions of the peaks were used as the sampling correction. The weights are inversely proportional to the square of the widths of these distributions. The derived corrections were then used to refit the showers, and the procedure was repeated until there was no appreciable change in the parameters.

The shower core is not well determined because Milagrito is smaller than the lateral size of a shower and the relatively thin water layer above the PMTs leads to large fluctuations in the observed pulse heights. As a result, shower curvature was not evident in the data. Nevertheless, Monte Carlo studies indicate that the angular resolution improves if a fixed, pulse-height-independent curvature of 0.04 ns/m is applied. Consequently this curvature correction was used in the shower fits. Figures 11 and 12 show the sampling correction and the RMS timing resolution as functions of the pulse height, respectively. The timing difference between the measured PMT times and the shower plane for several ranges of pulse height are shown in Figure 13. Note that the distributions are highly non-Gaussian at low light levels with long "late-light" tails.

Selection criteria can be derived for the hit PMTs to minimize the problems associated with a $\chi^2$ fitter. The selection criteria were determined by studying the quantity $\Delta_{EO}$, which is related to the angular resolution. $\Delta_{EO}$ is obtained by fitting each shower with two independent, interleaved portions of the detector (the detector is divided as light and dark squares of a checkerboard) and computing the difference in the fit space angles. It should be noted that this technique is not sensitive to certain systematic errors, such as those due to core location errors. In the absence of these systematic effects, $\Delta_{EO}$ should be about twice the overall angular resolution [15]. By minimizing the $\Delta_{EO}$ distribution the following procedure was developed to optimize the $\chi^2$ fitter. Figure 13 shows that the timing distributions become more Gaussian for larger pulse heights. For the first iteration, all PMTs with more than 2 PEs were used in the fit. This cut removed the most non-Gaussian distributions from consideration. In three subsequent iterations, only those PMTs whose contribution to the $\chi^2$ was less than 9, 6.25, and 4, respectively, or which were early with respect to the previous fit, were used to refit the shower direction. Figure 14 shows the resulting mean value of $\Delta_{EO}$ vs. the number of PMTs used in the fit, $N_{Fit}$, for data taken with a water depth of 2 m. Roughly 90% of all triggers (with a hardware requirement of at least 95 struck PMTs) are successfully fit with this algorithm. The remaining 10% of the triggers are believed to be associated with single muons passing through the detector at nearly horizontal angles (~6%) and single (unaccompanied) hadrons (~3%) [16]

Figure 14 shows that the angular resolution for Milagrito improves with increasing $N_{Fit}$. Figure 15 compares the observed median of the $\Delta_{EO}$ distribution vs. $N_{Fit}$ for 1 m of water with the expectation from a Monte Carlo simulation. Figure 16 shows the $N_{Fit}$ distribution for the three different water depths. Figure 17 compares the $\Delta_{EO}$ distributions with $N_{Fit} \geq 40$ for the three different water depths. There is a marked improvement between depths of 1 m and 1.5 m. The optimal water depth is a function of the $N_{Fit}$ distribution, the distributions of $\Delta_{EO}$ as a function of NFit and the event rate. The procedure used to determine the optimal water depth is described below. To simplify the analysis the significance of a binned source search was maximized.

Because the point-spread function for Milagrito is non-Gaussian, the standard formulae for calculating the optimal bin size to use in a search for a signal in the presence of background are not appropriate. The best cut on $N_{Fit}$ and the optimal bin size can be derived assuming that $\Delta_{EO}$ is twice the point-spread function and that the distributions of $N_{Fit}$ are the same for proton-induced and photon-induced showers. This was done by constructing $\Delta_{EO}$ distributions for bands of $N_{Fit}$ and using these together with the observed $N_{Fit}$ distribution to simulate the distribution of signal events from a point source. Figure 18 shows the optimal bin size as a function of the expected number of background events in 2° x 2° bin on the sky for data with 1 m of water. The figure also indicates the best cut on $N_{Fit}$ as a function of the expected number of background events in this bin. Using the above results, a signal observed with a significance of n$\sigma$ in Milagrito with a water depth of 1 m, would be observed with a significance of 1.2 n$\sigma$ (1.25 n$\sigma$) with a water depth of 1.5 m (2 m) due to the increased event rate and the improved angular resolution.

As stated above, $\Delta_{EO}$ gives no information on systematic pointing errors. In fact, Milagrito has a systematic pointing error caused by the late tail in the timing resolution (Fig. 13). Some of this late light was caused by shower particles that lagged the shower front and some from light that scattered in the detector. However, most of the late light was caused by nearly horizontal light from the Cherenkov cone traveling long distances before striking a PMT. This late light leads to a systematic error: the shower plane is reconstructed at a larger zenith angle than the incident particle. (To reduce the impact of this late light in Milagro, reflecting baffles were installed on each PMT to block out horizontal light. The baffles also increase the effective light collection of each PMT.) Monte Carlo studies show that this systematic error is expected to be approximately proportional to the zenith angle and is about $1^o$ at a zenith angle of $45^o$. Studies of the apparent displacement of the shadow of the moon as a function of zenith qualitatively confirm this effect (see below) [17].

**8. Operation of Milagrito**
The operation of Milagrito began on February 8, 1997 and ended on May 7, 1998. Most of the data were taken with 1 m, 1.5 m, or 2 m water above the tops of the PMTs. Data were acquired continuously except for periods of construction as noted below.

Milagrito operated largely unattended. During normal operations, the experiment was periodically monitored. An automatic alert was generated under serious error conditions, such as loss of electrical power, an abnormal event rate (including zero), or an abnormal water pressure in the recirculation system. The nature of the error condition, the time of day, and road conditions determined the response time and the subsequent restart of data taking. Many of these problems resulted from power outages, which were often weather related (snow or high winds); in these cases, it made sense to wait for the power to stabilize before traveling to restart the experiment. Less serious errors could be corrected remotely. For example, individual high-voltage channels could be turned on or off, and the data-taking computers could be rebooted remotely. There were also scheduled downtimes to accommodate construction activities, such as the installation of the lightning protection system.

Water or snow on top of the cover led to significant decreases in the trigger rate. This effect is similar to increasing the atmospheric overburden above the detector. It should be noted that the sloping sides of the pond amplified the depth of collected water or snow on top of Milagrito by a factor of three; the pond is full during operation of Milagro so this amplification does not occur.

The average daily Milagrito trigger rate over the period of operation is shown in Figure 19. The decrease in event rate at around MJD50510 was caused by an increasing amount of snow and water on top of the cover; this water was pumped off the cover starting ~MJD50530. Installation of the lightning protection system caused many of the outages around MJD50680. The increase in event rate starting at MJD50790 (50840) was caused by adding water inside the pond to increase the water level above the PMTs to 1.5 m (2 m). The increase in event rate around MJD50890 was the result of pumping the snowmelt off the top of the pond cover. The final drop in event rate was due to continued data taking while the pond was emptied. In total, $8.9 \times 10^9$ events were acquired and written to tape. Of these, $5.3 \times 10^9$ events were taken with a water depth of 1 m, $1.1 \times 10^9$ at 1.5 m, and $2.5 \times 10^9$ at 2 m.

The Milagrito detector was live about 79.5% of the available time during the 15 months it operated. Most of the downtime was attributable to power outages (~11.5%), calibrations (~3%), and scheduled maintenance and construction activities (~3%). The remaining downtime was due to hardware or software errors, many of which have been remedied.

Over the lifetime of Milagrito, 34 PMTs stopped working and had to be turned off. Examination of these tubes when Milagrito was dismantled indicated that most of these problems were caused by the breakdown of some of the wire-wound resistors in the PMT base. There was no evidence of water leaking into the encapsulated PVC housing. The PMT bases for Milagro were built with carbon composition resistors to avoid this problem.

**9. Milagrito Sensitivity**
The response of Milagrito to cosmic rays and gamma rays was simulated in two steps: the cascade in the atmosphere down to the local ground level, and the detector response including the particle interactions in the water as well as the production of Cherenkov photons and their detection by the photomultiplier tubes. The model of the atmosphere is based on the US standard atmosphere. It uses a 5-layer parameterization with the lower four layers having an exponential density vs. altitude dependence and the uppermost layer following a linear density vs. altitude dependence. The air shower cascade in the atmosphere uses CORSIKA (version 5.61) [18], a simulation package provided by the KASCADE group [19]. CORSIKA uses EGS 4 [20] for electromagnetic interactions, while several different options are available for hadronic interactions. The GHEISHA [21] code is used for low energy hadronic interactions ($E \leq 80$ GeV) and VENUS [22] is used for high-energy hadronic interactions, including the first interaction. The incident cosmic ray fluxes for protons, helium, and CNO primaries were taken from references 23 and 24. Gamma rays were generally thrown with an $E^{-2.5}$ spectrum.

The simulation of the Milagrito detector is based on the GEANT package (version 3.21) [25]. GEANT includes electromagnetic and hadronic interactions and the production of Cherenkov photons. The explicit production and tracking of δ-rays is important because of their ability to produce Cherenkov light; they contribute ~10% of the total number of photoelectrons per event. GEANT explicitly generates and tracks Cherenkov photons with the correct spectrum. Thus the wavelength dependence of the absorption and the refractive index of water, the reflectivity of the materials in the pond, and the quantum efficiency of the PMTs can be taken into account. A comparison between experimental and simulated data shows that the absorption and scattering lengths of the water and the reflectivity of the pond bottom and the cover material are important variables in the simulation. The Milagrito simulation assumes isotropic scattering of Cherenkov photons from the pond bottom. Some fraction of the cover is not in direct contact with the water due to pockets of trapped air, so that total internal reflection significantly increases the reflectivity at the top of the water. It is difficult to estimate the area over which total internal reflection takes place, especially as it changed with time. Thus the two extremes, no total internal reflection and complete internal reflection, were simulated.

The difference between the measured PMT time and the shower plane, $\chi_t$, for different numbers of detected photoelectrons is strongly influenced by the amount of light scattering, absorption and reflection; this ultimately affects the angular resolution of the detector. Figure 13 shows the $\chi_t$ distribution for PMTs as a function of the number of detected photoelectrons, for experimental data (dotted line) and simulated data (solid line). Results presented here are from the simulation assume a 20-m absorption length above 400 nm and a 20-m scattering length in water, 5% reflectivity of the pond bottom, and complete internal reflection at the water-air-cover boundary. Large negative values of $\chi_t$ indicate late light. The late light tail at low light levels, discussed previously, is qualitatively described by the simulation.

The full simulation of the air shower and the detector response allows various detector properties to be estimated, such as the effective area, $A_{eff}$, for both photon- and cosmic-ray-induced air showers as a function of the energy and incident zenith angle of the primary particle. If the simulated showers are thrown with cores randomly distributed over a sufficiently large area, $A_{core}$, the effective area is defined by:
$$A_{eff} = (n_{trigger} / n_{total}) \times A_{core},$$
where $n_{total}$ is the total number of simulated showers and $n_{trigger}$ is the number of showers fulfilling the trigger and reconstruction conditions.

Figure 20a shows $A_{eff}$ for photon showers and cosmic-ray showers initiated by protons, helium and nitrogen, vs. the energy of the primary particle. Figure 20b shows how $A_{eff}$ varies with zenith angle for photon showers. For these figures, showers were thrown with zenith angles between $0°$ and $45°$ and energies from 100 GeV and 500 TeV for photon- and proton-induced air showers, between 400 GeV and 500 TeV for showers initiated by helium, and between 1.4 TeV to 500 TeV for showers initiated by nitrogen. About 14% of triggers have zenith angles above $45°$ in the data.

The effective area for proton-initiated showers is larger than that for photon-initiated showers below ~2 TeV. At energies above ~5 TeV, the larger effective area for photon-induced showers provides an intrinsic cosmic-ray background suppression.

The effective area is larger than the physical area of the pond above ~3 TeV for both photon- and proton-initiated showers. Only 21% of the photon-induced and 16% of cosmic-ray-induced showers triggering Milagrito have cores within the area of the pond. This leads to a rather broad energy distribution for detected events with no well-defined energy threshold. The median energy for triggered photon events varies slowly with source declination ranging from ~3 TeV for sources with declination $40°$ to 7 TeV for sources with declination about $20°$.

Showers produced by relatively low-energy gamma rays may produce too few particles at ground level to satisfy the Milagrito trigger requirement, but may nevertheless produce some particles that can be detected by the PMTs. A strong burst of low-energy particles may be recognized by an increase in the measured scaler rates. Figure 21 shows the effective area for the low-threshold scalers, $A_{low}$, for photons incident at a fixed zenith angle of $22°$ as a function of the energy of the primary photon. The total number of counts observed in the sum of the low-threshold scalers is the product of the incident flux, $A_{low}$, and the observation time. An increase in the scaler rate may also be an effective way to detect protons from a solar coronal mass ejection [26].

As in more conventional EAS arrays, the energy of the primary particle is inferred from a measurement of the particle distribution on the ground. The energy of the primary particle is related to the number of detected electromagnetic particles, the depth of the atmosphere traversed by the air shower, and the location of the shower core. For a fixed primary energy, the number of particles that reach the ground is a strong function of the depth in the atmosphere of the first interaction and is thus subject to wide fluctuations. The typical energy resolution of an EAS array is ~50%.

In a water Cherenkov detector such as Milagrito, there are also large variations in the amount of light collected by the PMTs from a single charged particle. This is caused by the different penetration depths of the particles in the water and the varying light-collection efficiencies for different particle trajectories relative to the PMT positions. Thus the number of struck PMTs in Milagrito is a better measure of the number of shower particles in the pond than the sum of the PMT pulse heights. The large fraction of events with cores outside the pond leads to a very poor determination of the core position and of the primary energy. Studies indicate that if the core position were determined to within 20 m, the energy resolution would be ~60%.

A good test of the simulation is a comparison between the predicted and measured event rates as a function of the zenith angle. Figure 22 shows the measured and simulated zenith-angle distributions for reconstructed events. With 1.0 m of water above the PMTs, the total rate predicted by the simulation agrees with the measured rate of ~330 Hz to within 20%. In the simulation, He and CNO contribute 21% and 4% of the total event rate, respectively. A fit of measured zenith-angle distribution to the form $\cos^n\theta$ yields n = 4.2, considerably flatter than EAS arrays using scintillation counters: the zenith-angle distribution for the CYGNUS experiment was consistent with n = 8.6 [27].

## 10. Milagrito Performance

Milagrito operated in a reliable and stable manner over a 15 month period. The event rate, 300–400 Hz, was chosen as a compromise between sensitivity and the cost of data storage. Figure 8 shows the trigger rate vs. number of PMTs required in coincidence for a water depth of 1 meter. The sharp increase in trigger rate below a ~50-tube requirement is believed to be due to triggers caused by single large-zenith-angle muons.

The cosmic-ray shadows of the sun and the moon have been used to measure the angular resolution of an EAS array above 50 TeV [28]. At TeV energies, the geomagnetic field should displace the moon's cosmic ray shadow [29] but it should still be observable. Figure 23 shows the shadow of the moon as observed by Milagrito. The displacement of the shadow relative to the actual position of the moon is due to the bending of the cosmic rays in the geomagnetic field (primarily a shift in right ascension) and the systematic pointing error caused by late light in Milagrito (primarily a shift in declination). The magnetic field of the sun also bends cosmic rays at TeV energies; this effect has been observed at somewhat higher energies by the Tibet air-shower array [30]. Figure 24 shows the shadow of the sun as observed by Milagrito; this shadow is obviously more dispersed than the shadow of the moon. An analysis of the shapes and locations of the shadow will be given elsewhere [17].

Markarian 501, a blazar, was the brightest known TeV source in the sky during 1997 [31, 32]. An analysis of the event excess from the vicinity of Markarian 501 observed with Milagrito yields a significance of 3.7$\sigma$ [33]. This is interpreted as a confirmation of Markarian 501 as a TeV source during this period and is indicative of the sensitivity of Milagrito. Results of a more complete analysis of observations of Markarian 501 with Milagrito has been published elsewhere [33].

**11. Milagro**
Milagrito was constructed primarily as a prototype for Milagro. Nevertheless, it was a fully functioning gamma-ray detector that collected data over 15 months. This paper has described the design, operation, and some early results from the Milagrito detector.

The water was drained from the pond and the Milagrito PMTs were removed in early summer, 1998. Milagro was installed in the fall of 1998, and engineering runs began in February, 1999. Milagro has 21,000,000 liters of water and two planar layers of PMTs in the pond. The top layer, has 450 PMTs on a 2.8 m x 2.8 m grid, 1.5 m below the water surface. This shower layer is similar to, but larger than Milagrito. The muon layer has 273 PMTs located under ~6 m of water, also on a 2.8 m x 2.8 m grid. The muon layer is used to measure the lateral distribution of the energy deposited in the pond by the shower particles: this information can be used to distinguish photon-initiated showers from hadron-initiated showers. Each PMT in Milagro has a reflecting conical baffle to increase its light-collection area and to prevent light traveling horizontally or upwards from reaching it. An array of 180 water detectors, each ~4.5 $m^2$ in area and 1-m deep, will be deployed over 10,000 $m^2$ surrounding the pond to help determine the location of the shower core.

Many valuable lessons were learned from Milagrito that will improve the performance of Milagro. Some of these have been described above. In addition, several minor improvements in the water recirculation system, especially making sure that the filters sealed in their housings, has resulted in water with improved attenuation lengths, as can be seen in Figure 3. Data taking with Milagro began on December 1, 1999. The sensitivity of Milagro is significantly better than that of Milagrito.

**ACKNOWLEDGMENTS**
We are indebted a great many people whose hard work helped see Milagrito reach fruition. In particular, we thank Steve Biller, Matteo Cavalli-Sforza, David Hale, Al Lu, Martin Adair, Mayra Alano, Emmanuel Bacolas, Michelle Beaver, Pieris Berreitter, Jeff Booth, Matthew Boyce, Tim Carlson, Richard Cho, David Cole, Kalle Cook, Stephen Cronin, Marcos Delgado, Joe DíIncennzo, Aaron Dinwoodie, Danial Elderege, Daniel Esquivel, Iana Fraser, Casey Gage, Enrique Gomez, Evan Graj, Joseph Herrera, Amy

Hodapp, Joseph Jun, Won Kim, Matt Knox, K. C. Kunes, Sabatay Lazar, Jay Lee, Chris Lopez, Lorena Lucille, J. Zane Macagnano, John MacMahon, Stephen Markacs, M. Masequesmay, Tracy Marsh, Alberto Martinez, Patrick McBride, Kurt McCormick, Ian McKenna, Matthew Millan, Davienne Monbleau, Jonathan Nix, Bryce Nordgren, Wansin Ounkeo, Phuc Phan, Mark Raugas, Raghu Reddy, MaryJane Sabihon, Brian Schwartz, Dave. R. Stegman, Duane Stemple, Jeremy Stoller, Shawn Wheelock, Danica Wyatt, Chadwick Young, and Andrew Zirm. We also thank the authors of CORSIKA for providing us with the simulation code and Rex Tayloe for help with implementing the detector simulations.

This work was supported in part by the National Science Foundation, The U. S. Department of Energy Office of High Energy Physics, The U. S. Department of Energy Office of Nuclear Physics, Los Alamos National Laboratory, the University of California, the Institute of Geophysics and Planetary Physics, The Research Corporation, and the California Space Institute.

**FIGURES**

Fig. 1: An event display from Milagrito. The length of each vertical line is proportional to the time at which that PMT was struck by light produced by particles in the shower front. This illustrates how the PMT signal times are used to reconstruct the incident direction of an air shower.

Fig. 2: An aerial photograph of the Milagrito site. The pond that housed Milagrito was empty when this photograph was taken. The utility building and electronics trailer are visible immediately to the left of the pond.

Fig. 3: The attenuation length of the water in the pond vs. date. The data for 1996 and 1997 are for Milagrito, while the data starting July, 1998 are for Milagro. The smaller errors for the last point on this figure are due to the use of a longer water cell (1 m rather than 20 cm) in the measuring device.

Fig. 4: The time response for single photoelectrons for a typical Hamamatsu R5912SEL photomultiplier operating at a gain of ~$10^7$.

Fig. 5: The measured ADC spectrum for a typical Hamamatsu R5912SEL photomultiplier showing the ADC pedestal (channel ~570) and the single PE peak (channel ~625). The conversion to charge is 50 fC/channel.

Fig. 6: The observed number of photoelectrons vs. the expected number for 5 Hamamatsu R5912SEL photomultipliers as measured with a pulsed laser and attenuating filters.

Fig. 7: A conceptual drawing of the dual TOT method illustrating the single logic pulse produced for a PMT pulse that triggers only the low threshold, and the two pulses separated by the high TOT for a pulse that triggers the high threshold.

Fig. 8: The trigger rate vs. the number of PMTs required in coincidence for a water depth of 1 m.

Fig. 9: A block diagram of the major components of the laser calibration system.

Fig. 10: The fractional error in the charge measurement as a function of charge (in units of PEs). The discontinuity near 3 PEs is due to the transition between the use of the low and high-threshold TOT measurements.

Fig. 11: The sampling correction as a function of the pulse height in a PMT.

Fig. 12: The RMS timing resolution as a function of the number of detected photoelectrons in a PMT.

Fig. 13: The difference between the shower plane and the measured PMT time, $\chi_t$, for PMTs with the number of detected photoelectrons in the ranges 0–1.5, 1.5–3, 3–5, and 5–10, for experimental data (dotted line) and simulated data (solid line). Note that $\chi_t$ increases as the PMT time gets earlier.

Fig. 14: The mean value of $\Delta_{EO}$ vs. the number of PMTs used in the fit, $N_{Fit}$, for data taken with a water depth of 2 m.

Fig. 15: The median value of $\Delta_{EO}$ vs. $N_{Fit}$ for data (with a water depth of 1 m) and Monte Carlo simulations.

Fig. 16: The distribution of the number of events vs. $N_{Fit}$ for water depths of 1 m (solid histogram), 1.5 m of water (dashed histogram), and 2 m of water (dotted histogram). The three histograms are normalized to have the same areas. Note that the number of working PMTs is higher for the 1-m data.

Fig. 17: The distribution of $\Delta_{EO}$ for events with $N_{Fit} \geq 40$ for the three different water depths. The dashed histogram is for 1 m, the solid is for 1.5 m, and the dotted is for 2 m. Note that the three histograms are normalized to have the same areas.

Fig. 18: The optimal bin size as a function of the expected number of background events in a 2° x 2° bin on the sky for data with 1 m of water. The figure also indicates the best cut on $N_{Fit}$ as a function of the expected number of background events in this bin.

Fig. 19: The average daily event rate over the operating life of Milagrito.

Fig. 20: (a) The effective area, $A_{eff}$, as a function of the energy of the primary particle for hadron and photon showers.

(b) The dependence of $A_{eff}$ with zenith angle for photon showers.

Fig. 21: The effective area of the Milagrito low-threshold scalers for cosmic gamma rays incident at a fixed zenith angle of 22° as a function of energy.

Fig. 22: The zenith angle distribution for data (histogram) and Monte Carlo simulation (points).

Fig. 23: The shadow of the moon as observed by Milagrito. The actual moon position is at the origin of the figure. The contours are in units of standard deviations ($\sigma$) of the event deficit.

Fig. 24: The shadow of the sun as observed by Milagrito. The actual sun position is at the origin of the figure. The contours are in units of standard deviations ($\sigma$) of the event deficit.

**REFERENCES**

[1] Ong, R., Physics Reports, **305**, 93 (1998).
[2] Hoffman, C. M., C. Sinnis, P. Fleury, and M. Punch, Rev. Mod. Physics **71**, 897 (1999).
[3] Fichtel, C. E. and J. L. Trombka, *Gamma-Ray Astrophysics*, NASA Reference Publication 1386, (1997).
[4] Weekes, T. C., *et al.*, in *Proc. XXV Int. Cosmic Ray Conf.*, **5**, 157, (1997) (Durban).

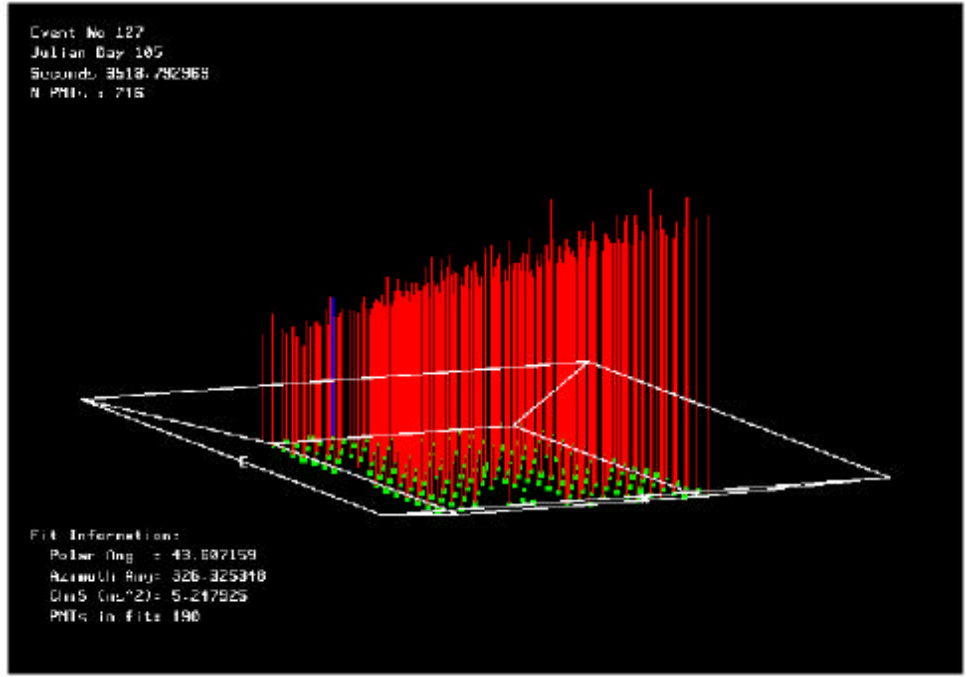

Fig. 1: An event display from Milagrito. The length of each vertical line is proportional to the time at which that PMT was struck by light produced by particles in the shower front. This illustrates how the PMT signal times are used to reconstruct the direction of an air shower.

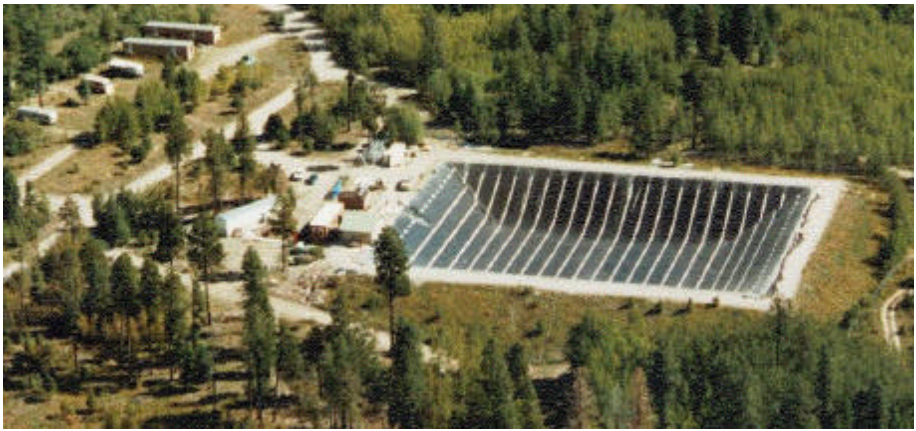

Fig. 2: Aerial photograph of the Milagrito site. The pond that housed Milagrito was empty when this photograph was taken. The utility building and electronics trailer are visible immediately to the left of the pond.

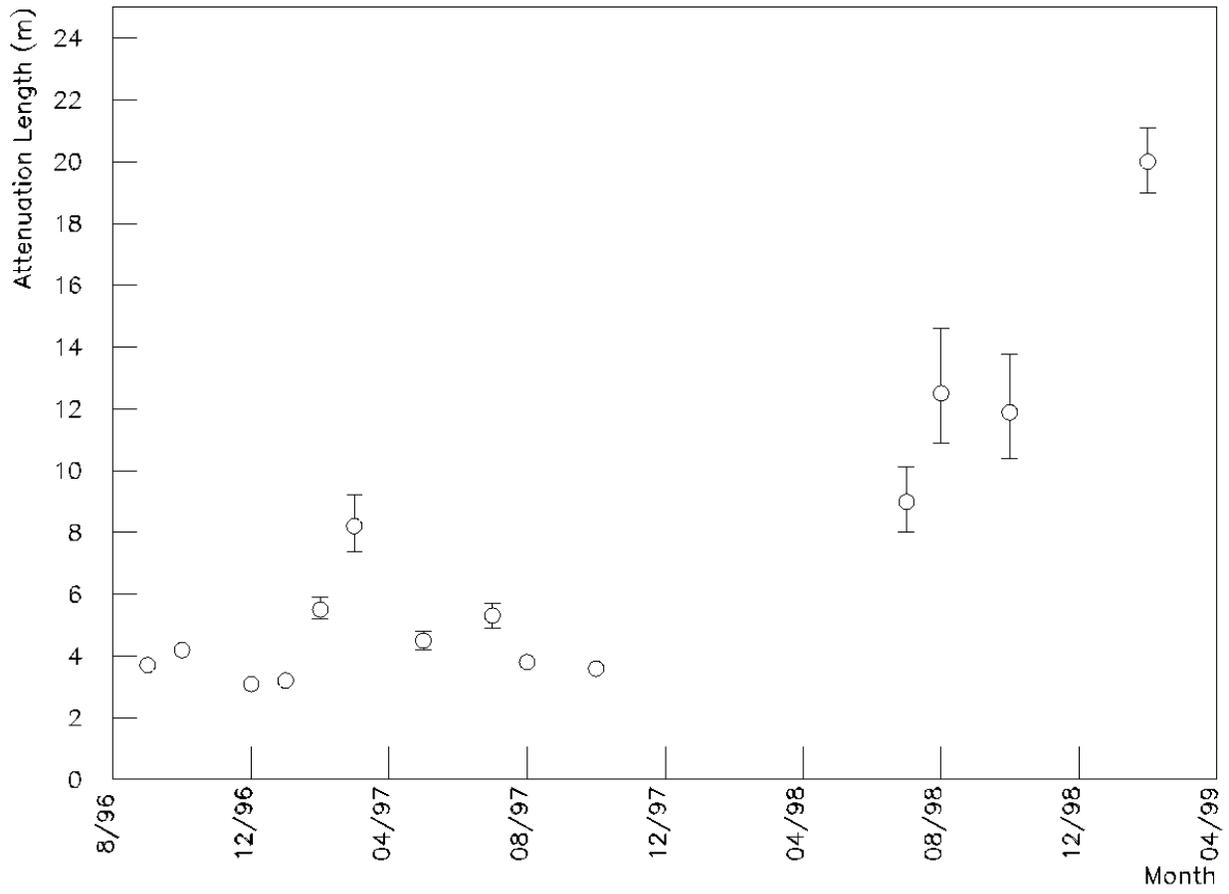

Fig. 3: The attenuation length vs. wavelength of the water in the Milagrito pond.

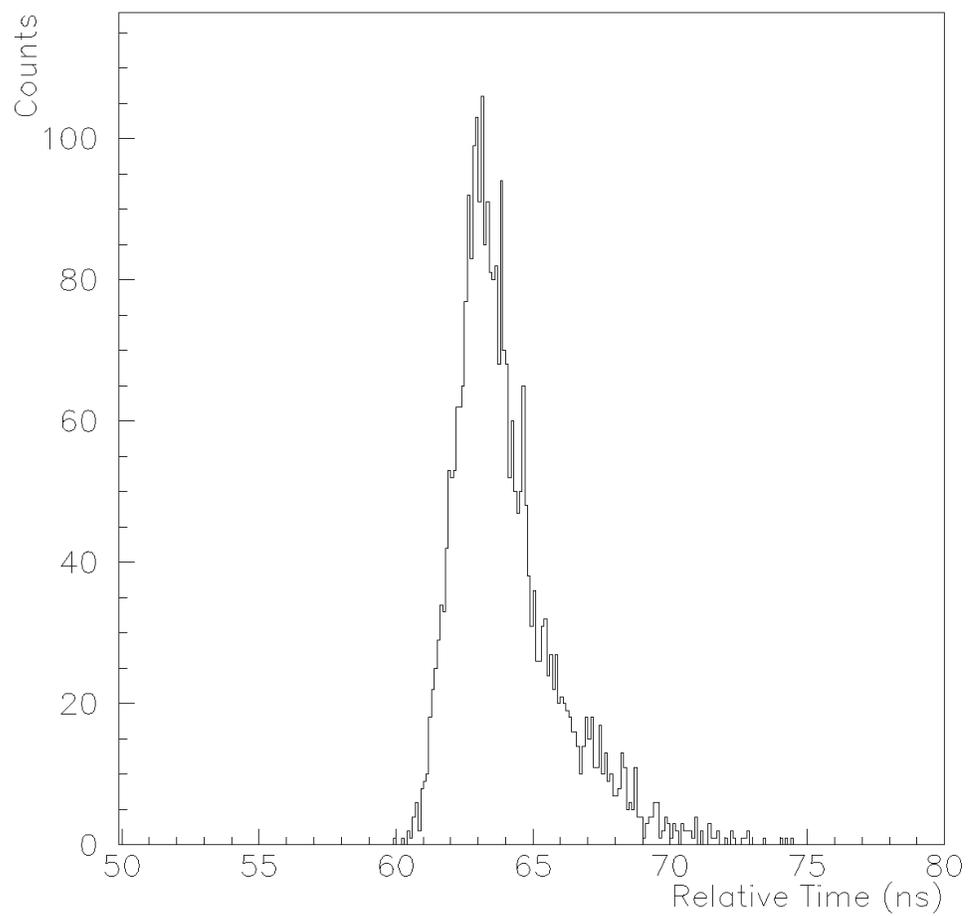

Fig. 4: The time response for single photoelectrons for a typical Hamamatsu R5912SEL photomultiplier operating at a gain of ~$10^7$.

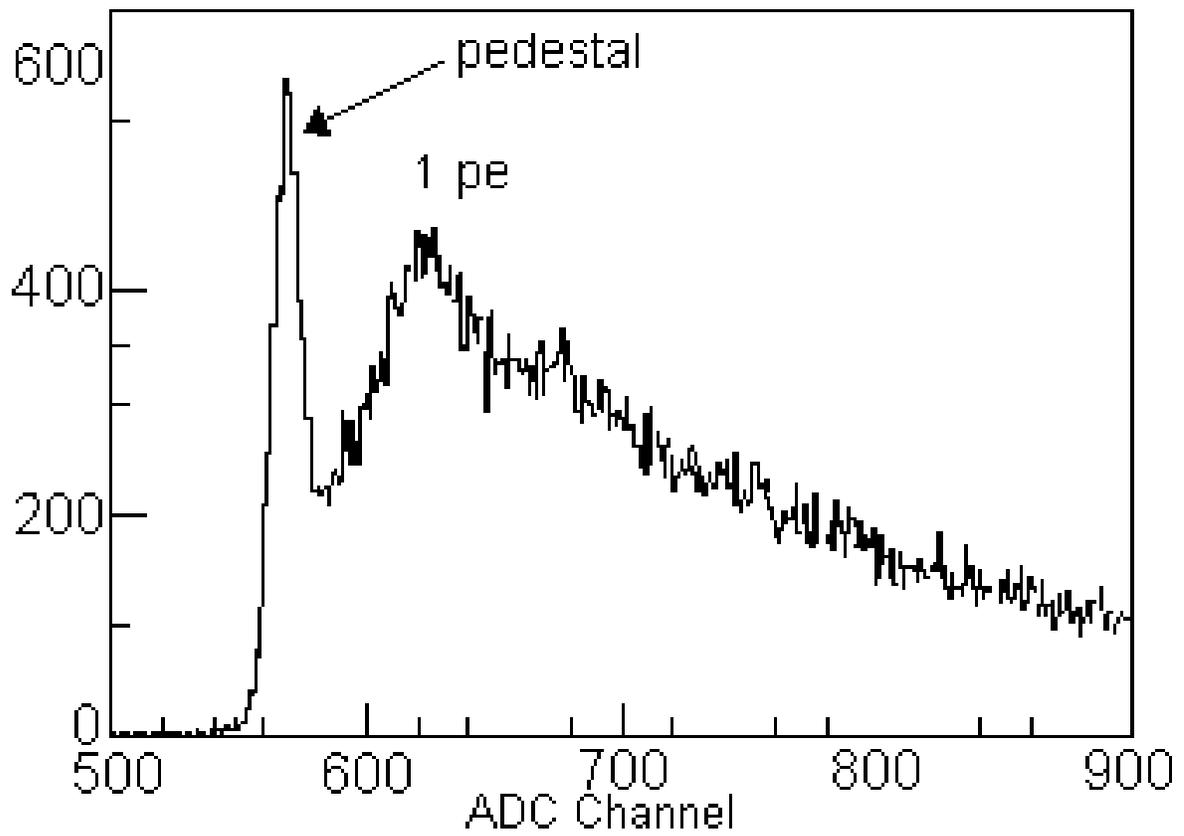

Fig. 5: The measured pulse charge spectrum for a typical Hamamatsu R5912SEL photomultiplier showing the ADC pedestal (channel ~570) and the single photoelectron peak (channel ~625). The conversion to charge is 50 fC/channel.

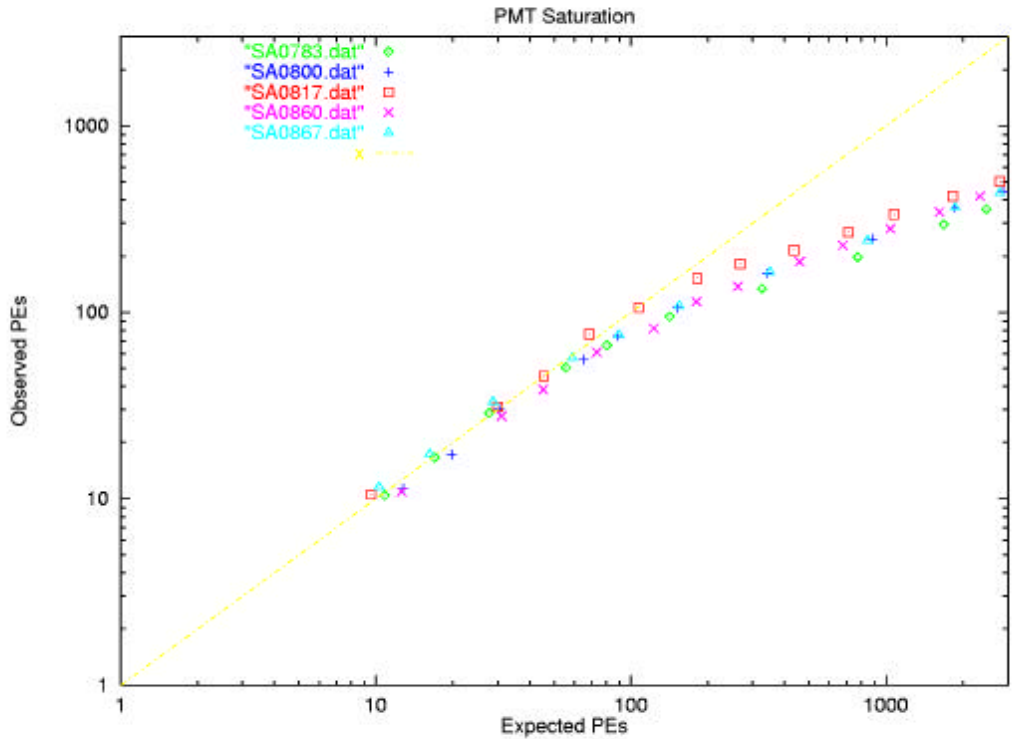

Fig. 6: The observed number of photoelectrons vs. the expected number for 5 Hamamatsu R5912SEL photomultipliers as measured with a pulsed laser and attenuating filters.

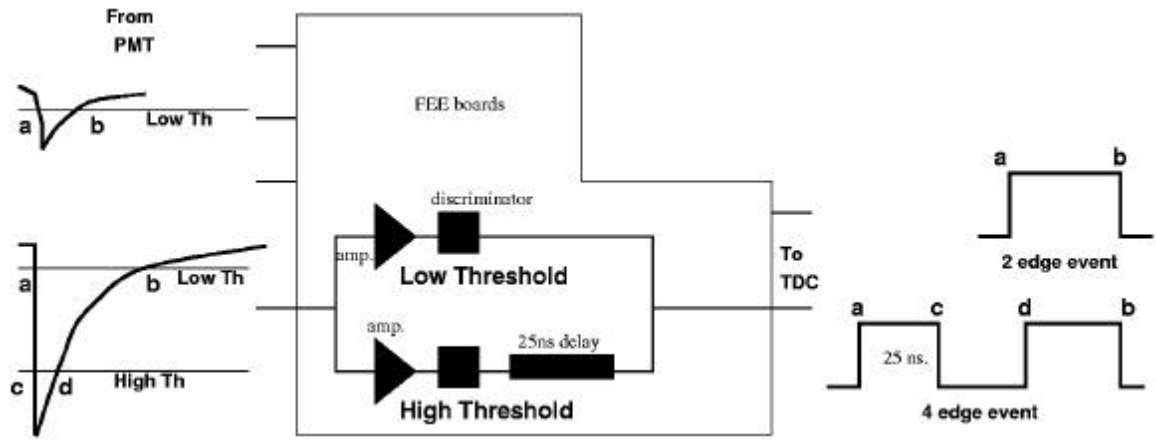

Fig. 7: A conceptual drawing of the dual TOT method illustrating the single logic pulse produced for a PMT pulse that triggers only the low threshold, and the two pulses separated by the high TOT for a pulse that triggers the high threshold.

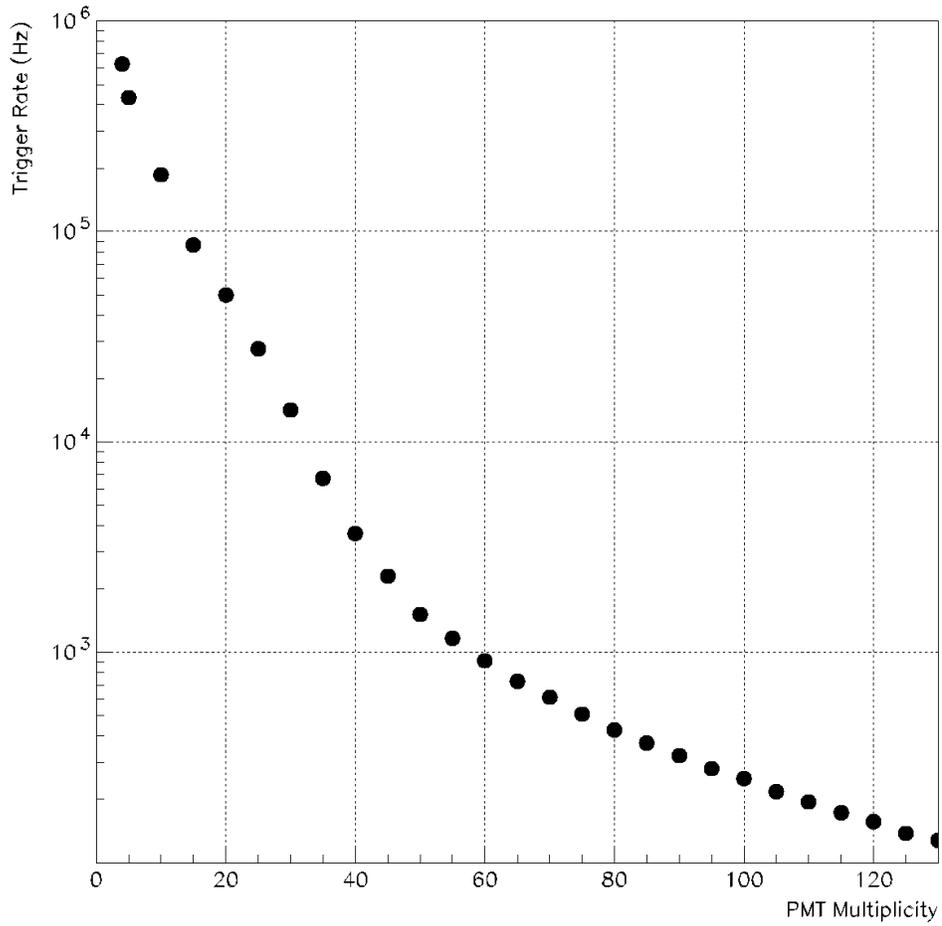

Fig. 8: The trigger rate vs. number of PMTs required in coincidence for a water depth of 1 m.

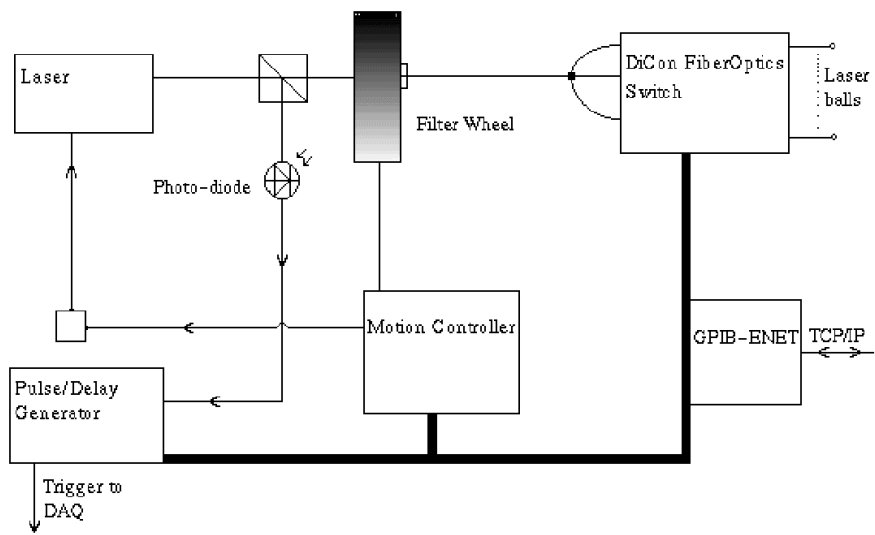

Fig. 9: A block diagram of the major components of the laser calibration system.

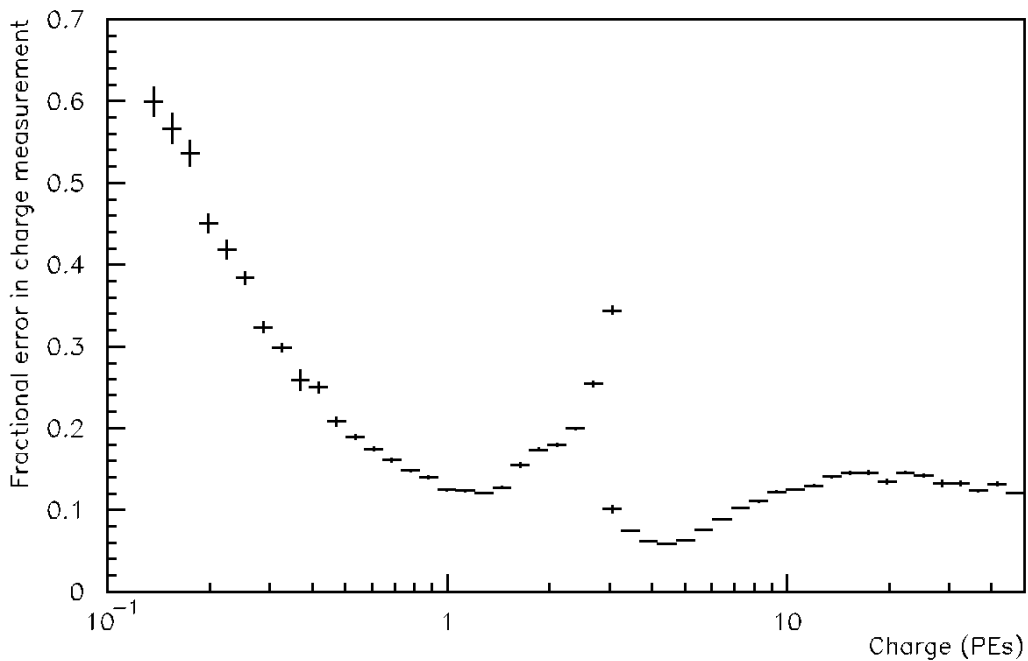

Fig. 10: The fractional error in the charge measurement as a function of charge (in units of PEs). The discontinuity near 3 PEs is due to the transition between the use of the low and high-threshold TOT measurements.

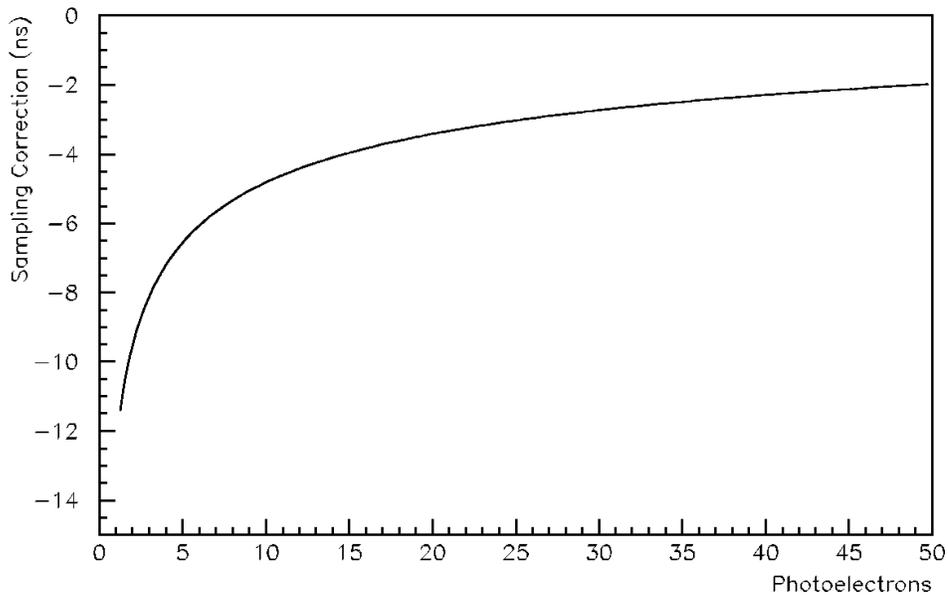

Fig. 11: The sampling correction as a function of the pulse height in a PMT.

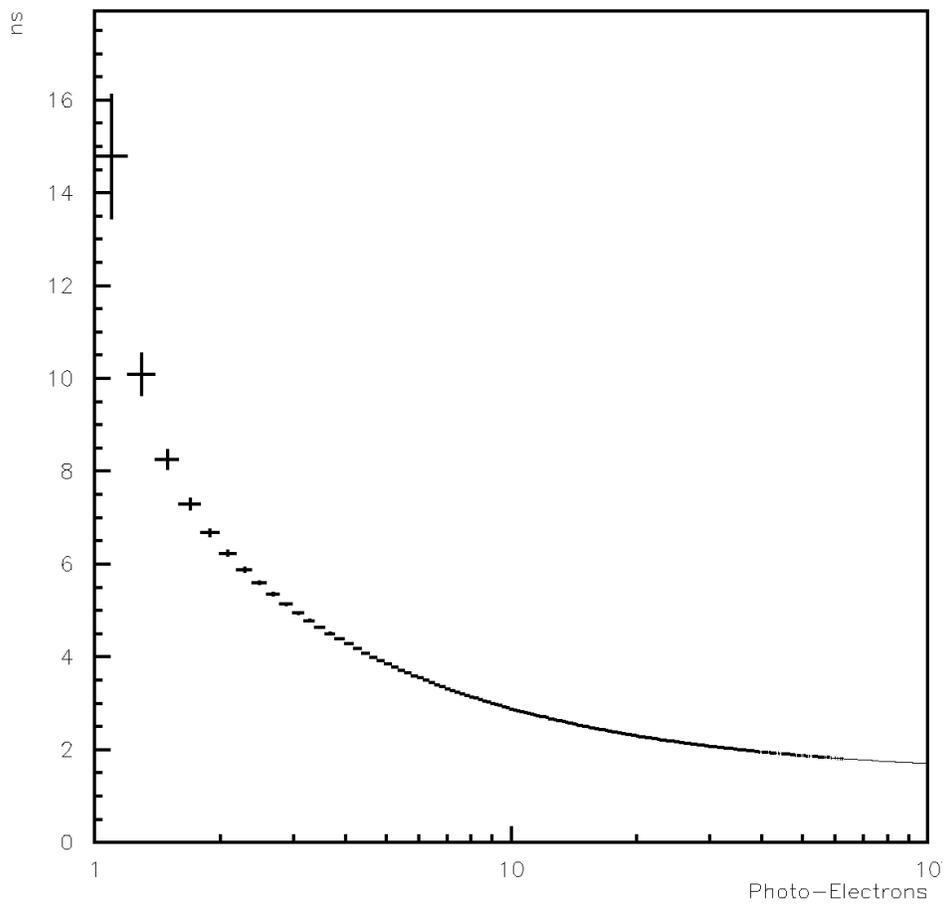

Fig. 12: The RMS timing resolution as a function of the number of detected photoelectrons in a PMT.

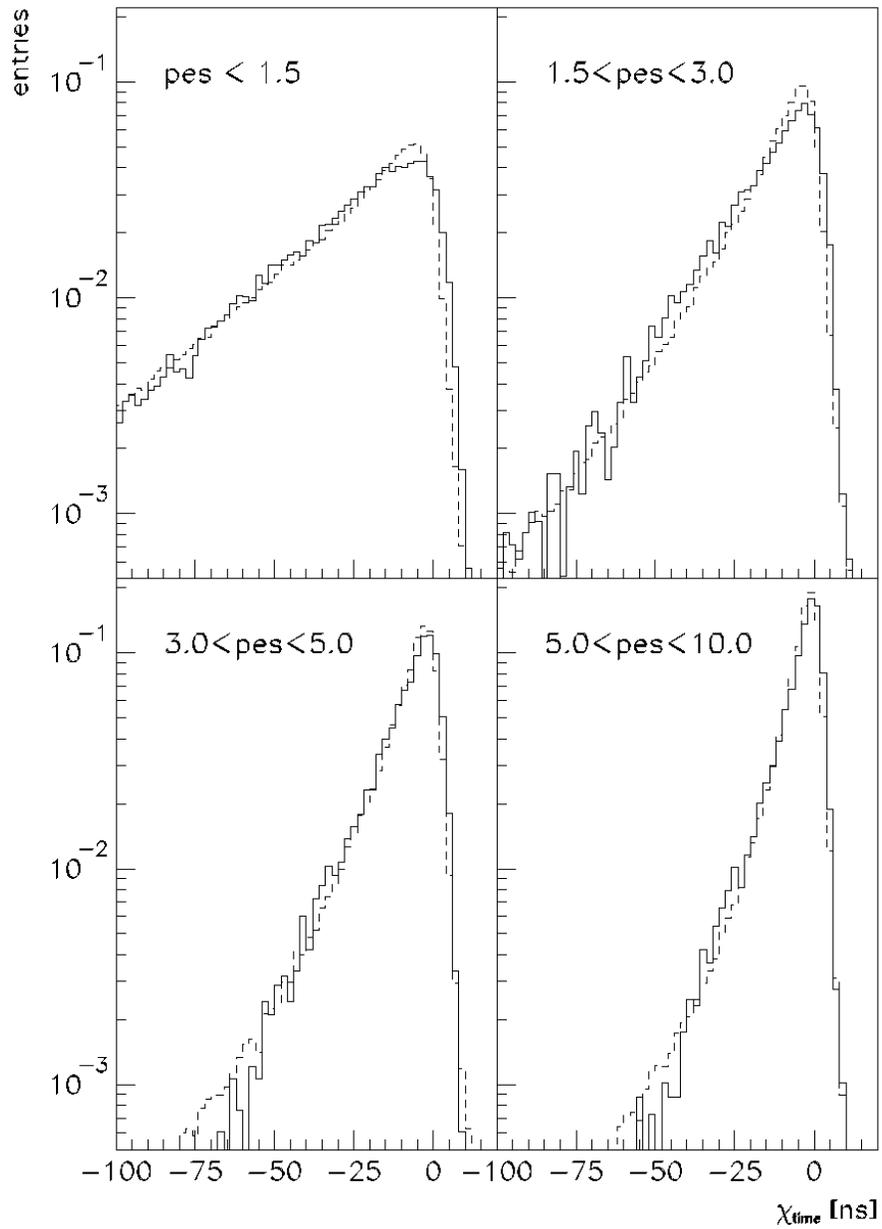

Fig. 13: The difference between the measured PMT time and the shower plane, $\chi_t$, for PMTs with the number of detected photoelectrons between 0 - 1.5, 1.5 -3, 3 - 5, and 5 - 10, for experimental data (dotted line) and simulated data (solid line). Note that $\chi_t$ increases as the PMT time gets earlier.

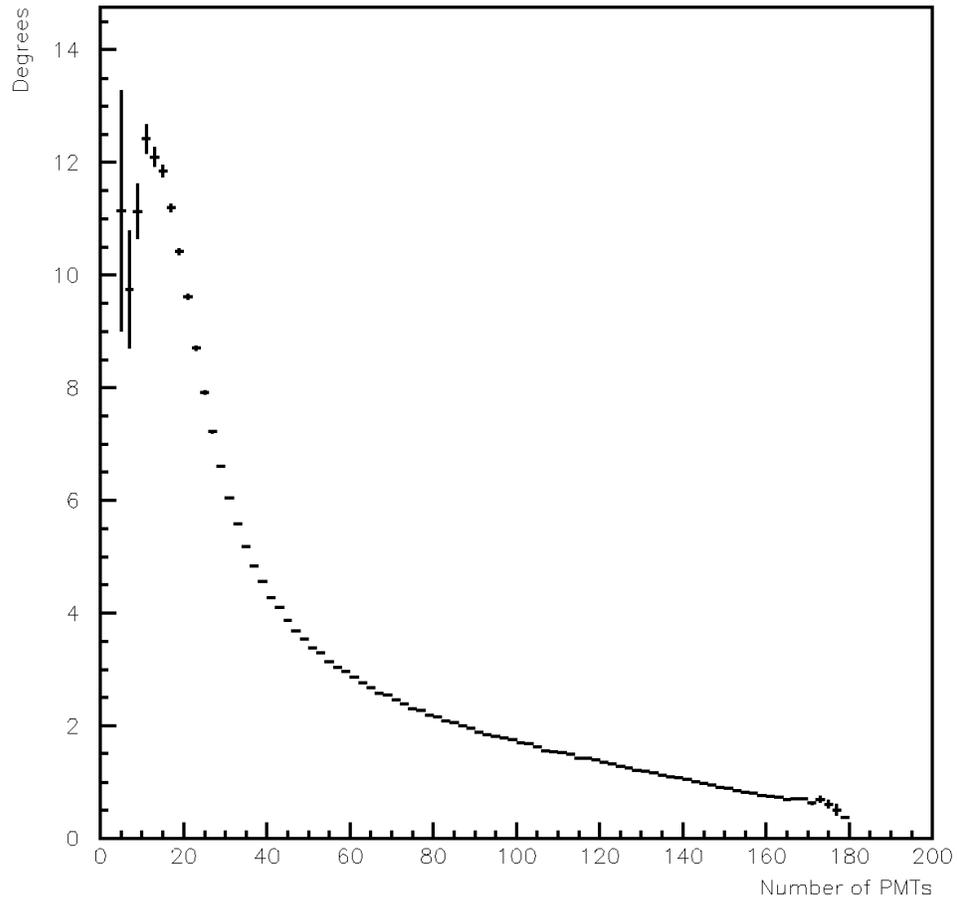

Fig. 14: The mean value of $\Delta_{EO}$ vs. the number of PMTs used in the fit, $N_{Fit}$, for data taken with a water depth of 2 m.

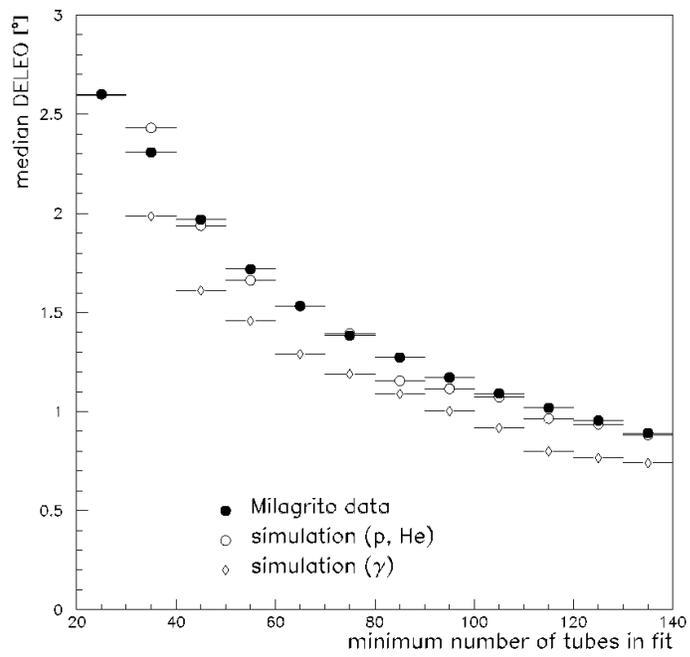

Fig. 15: The median value of $\Delta_{EO}$ vs. $N_{Fit}$ for data (with a water depth of 1 m) and Monte Carlo simulations.

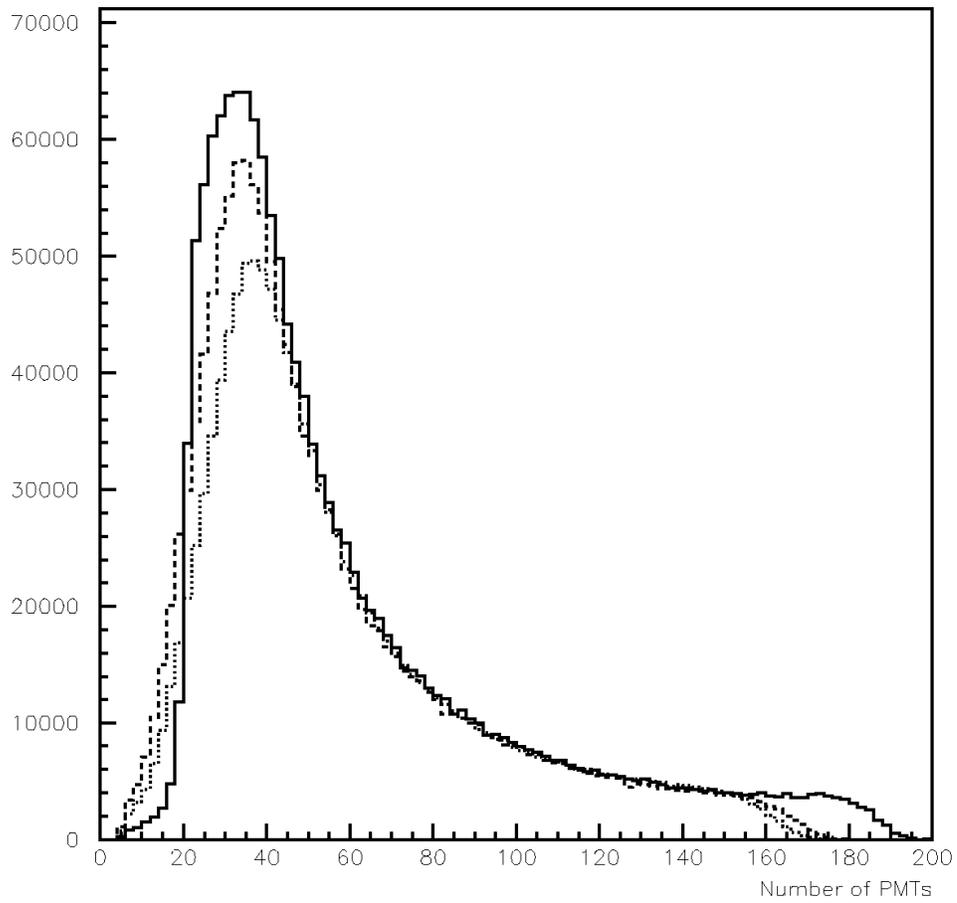

Fig. 16: The distribution of the number of events vs. $N_{Fit}$ for water depths of 1 m (solid histogram), 1.5 m of water (dashed histogram), and 2 m of water (dotted histogram). The three histograms are normalized to have the same areas. Note that the number of working PMTs is higher for the 1-m data.

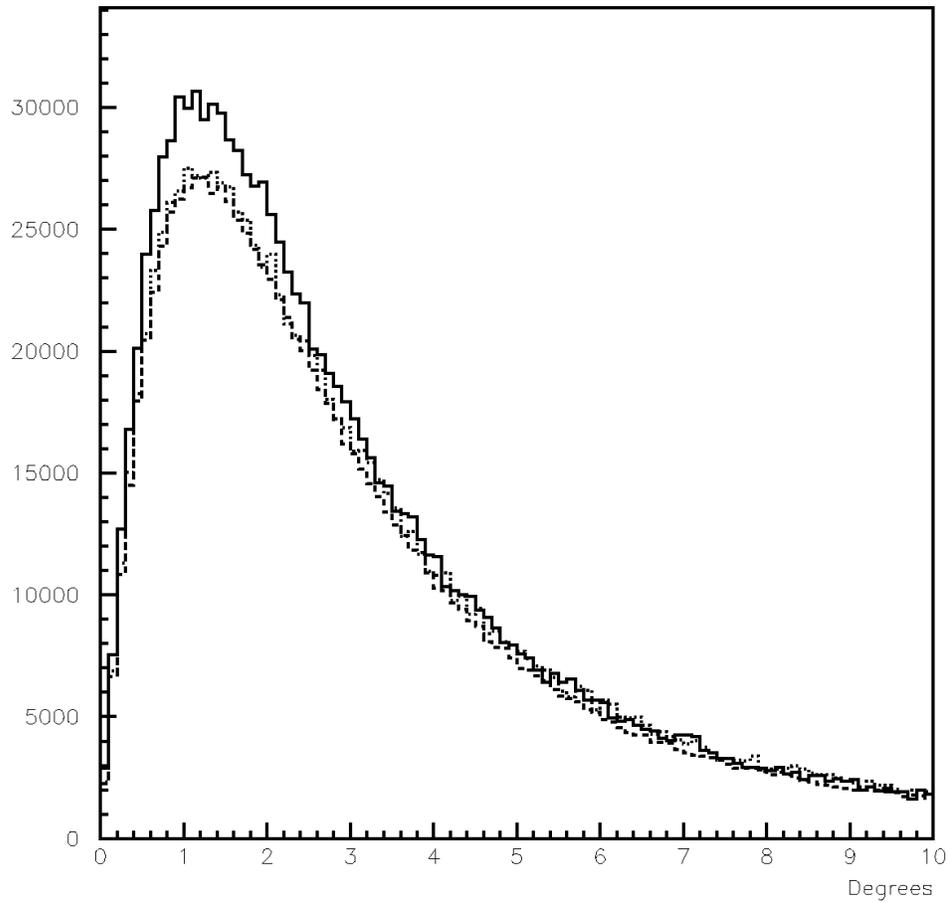

Fig. 17: The distribution of $\Delta_{EO}$ for events with $N_{Fit} \geq 40$ for the three different water depths. The dashed histogram is for 1 m, the solid is for 1.5 m, and the dotted is for 2 m. Note that the three histograms are normalized to have the same areas.

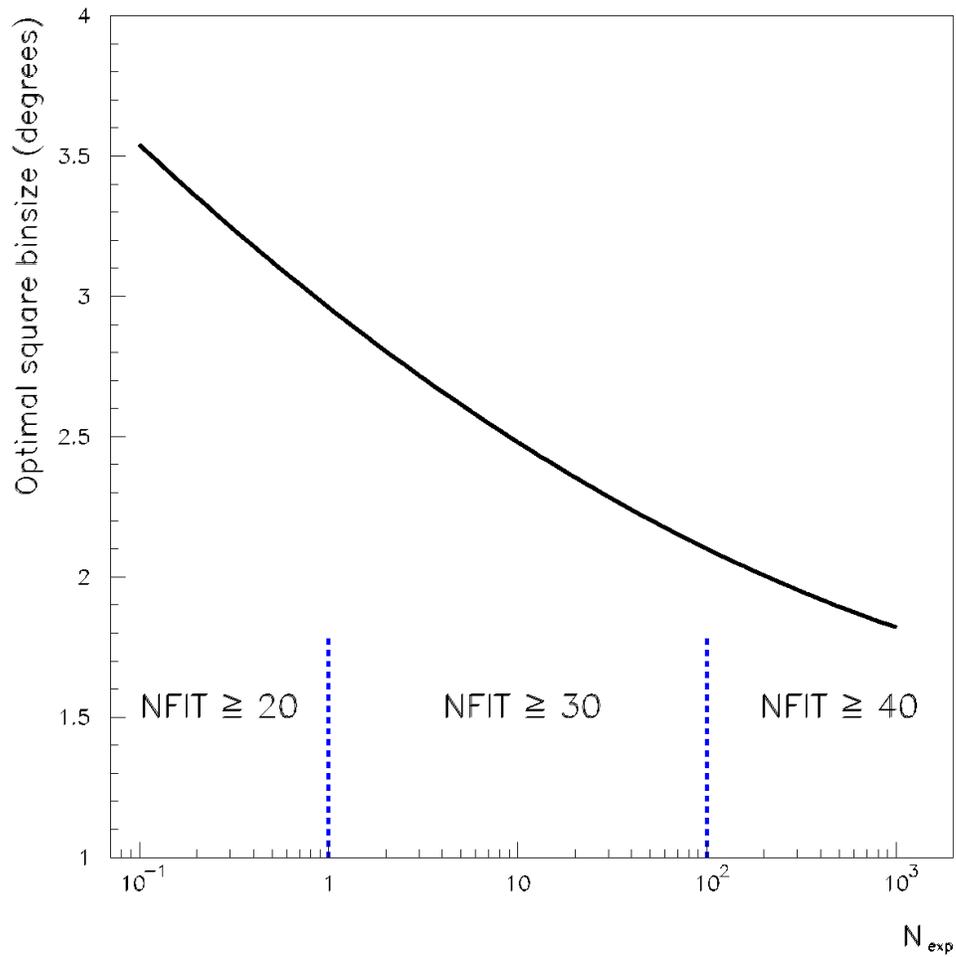

Fig. 18: The optimal bin size as a function of the expected number of background events in a 2° x 2° bin on the sky for data with 1 m of water. The figure also indicates the best cut on $N_{Fit}$ as a function of the expected number of background events in this bin.

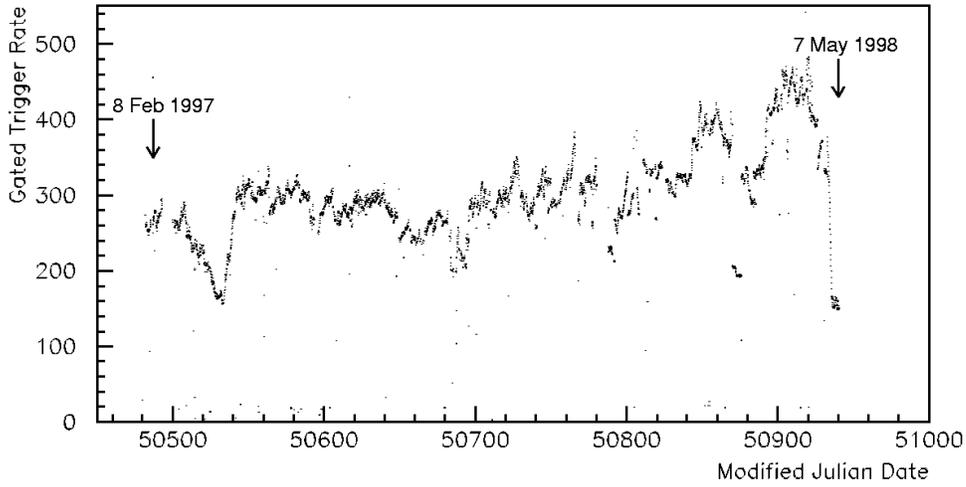

Fig. 19: The average daily event rate over the operating life of Milagrito.

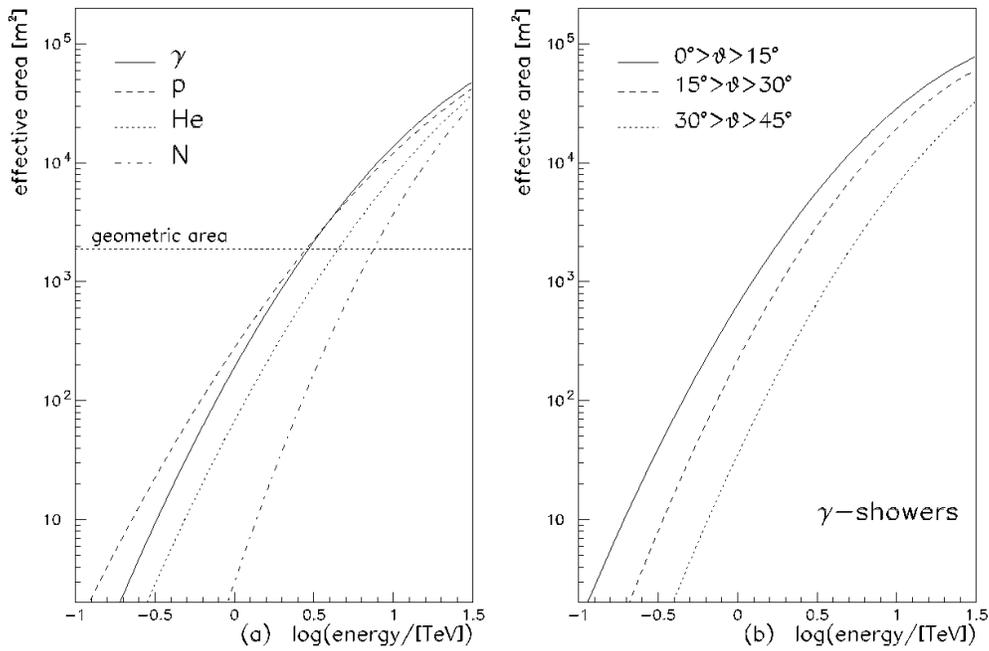

Fig. 20: (a) The effective area, $A_{eff}$, as a function of the energy of the primary particle for hadron and photon showers.
(b) The dependence of $A_{eff}$ with zenith angle for photon showers.

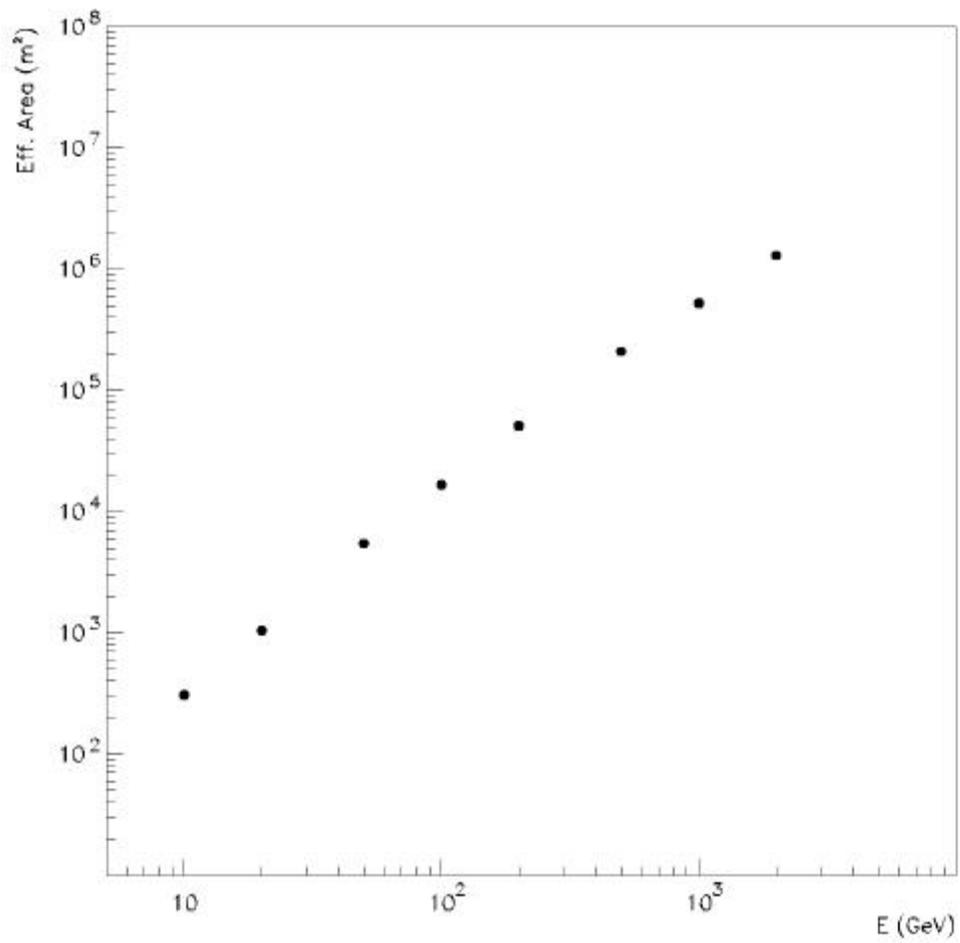

Fig. 21: The effective area as a function of energy for the scaler system. The squares are for cosmic gamma rays incident at a fixed zenith angle of 22° and the low-threshold scalers: the triangles are for isotropic cosmic protons and the high-threshold scalers.

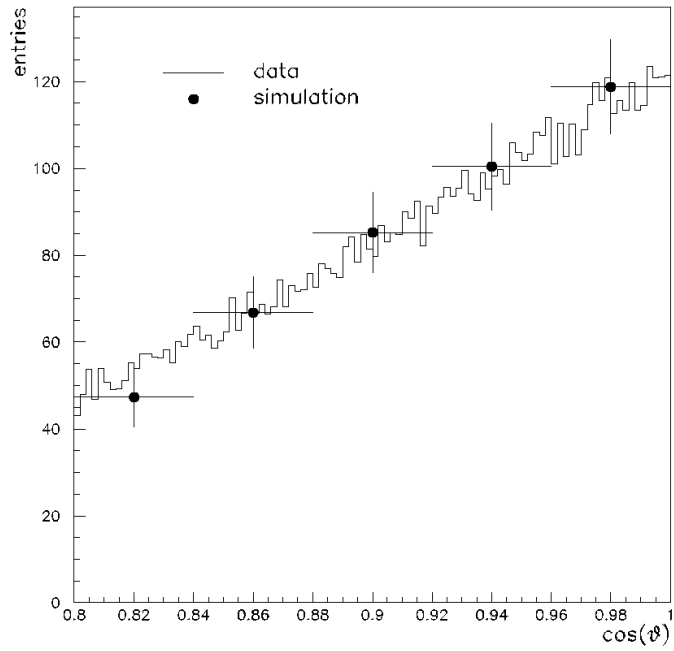

Fig. 22: The zenith angle distribution for data (histogram) and Monte Carlo simulation (points).

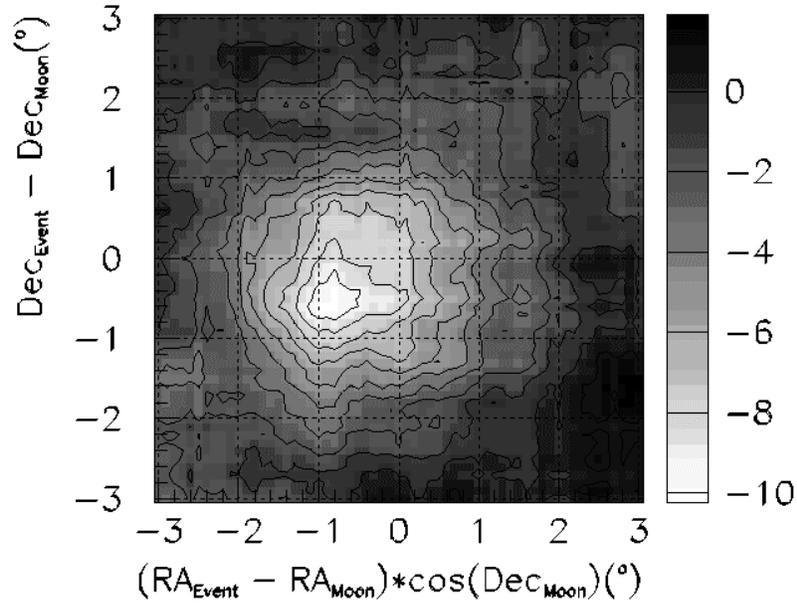

Fig. 23: The shadow of the moon as observed by Milagrito. The actual moon position is at the origin of the figure. The contours are in units of standard deviations (σ) of the event deficit.

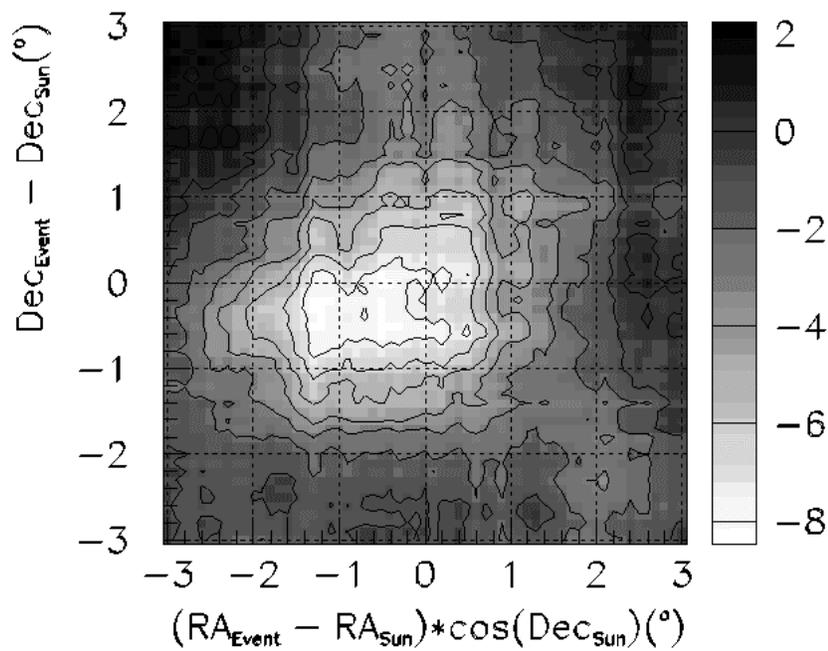

Fig. 24: The shadow of the sun as observed by Milagrito. The actual sun position is at the origin of the figure. The contours are in units of standard deviations (σ) of the event deficit.